%% file: HMP_astroph_v2.tex
\documentclass[usenatbib]{mn2e}
\usepackage{graphicx}

\input{journals}

\newcommand{\sm}[1]{\mbox{{\scriptsize #1}}}

\newcommand{\simge} {\,{}^>_{\sim}\,}

\def\ltsima{$\; \buildrel < \over \sim \;$}    
\def\lesssim{\lower.5ex\hbox{\ltsima}}           
\def\gtsima{$\; \buildrel > \over \sim \;$}    
\def\gtrsim{\lower.5ex\hbox{\gtsima}}           

\newcommand{\be}{\begin{equation}}
\newcommand{\ee}{\end{equation}}
\newcommand{\bea}{\begin{eqnarray}}
\newcommand{\eea}{\end{eqnarray}}
\newcommand{\bdm}{\begin{displaymath}}
\newcommand{\edm}{\end{displaymath}}
\newcommand{\bef}{\begin{figure}}
\newcommand{\eef}{\end{figure}}
\newcommand{\befone}{
  \begin{figure*}
  \centering
  \begin{minipage}{\textwidth}
  }
\newcommand{\eefone}{\end{minipage}\end{figure*}}

\newcommand{\cm}{\mbox{cm}}

\newcommand{\km}{\mbox{km}}
\newcommand{\kms}{{\rm ~km s}^{-1}}

\renewcommand{\sec}{\mbox{~s}}

\newcommand{\g}{\mbox{g}}
\newcommand{\gamef}{\gamma_{\rm e}}
\newcommand{\G}{\mbox{G}}
\newcommand{\Msol}{\mbox{$M_{\sun}$}}
\newcommand{\pc}{\mbox{pc}}
\newcommand{\pcc}{{\rm ~cm}^{-3}}

\newcommand{\K}{\mbox{K}}
\newcommand{\yr}{\mbox{yr}}

\newcommand{\Myr}{\mbox{Myr}}

\newcommand{\di}{\mbox{d}}

\newcommand{\vv}{{\bf v}}

\newcommand{\vB}{{\bf B}}

\def\fig@scaling{0.95}

\def\showone#1{
  \centering
  \leavevmode
  \includegraphics[width=\fig@scaling\linewidth]{#1.pdf}
}

\def\showonep#1{
  \centering
  \leavevmode
  \includegraphics[height=\fig@scaling\textheight]{#1.pdf}
}

\def\showtwover#1#2{
  \centering
  \leavevmode
  \includegraphics[width=\fig@scaling\linewidth]{#1.pdf} \hfil
  \includegraphics[width=\fig@scaling\linewidth]{#2.pdf}
}

\def\showthreeover#1#2#3{
  \centering
  \leavevmode
  \includegraphics[width=\fig@scaling\linewidth]{#1.pdf} \hfil
  \includegraphics[width=\fig@scaling\linewidth]{#2.pdf} \hfil
  \includegraphics[width=\fig@scaling\linewidth]{#3.pdf}
}

\def\showfourover#1#2#3#4{
  \centering
  \leavevmode
  \includegraphics[width=\fig@scaling\linewidth]{#1.pdf} \hfil
  \includegraphics[width=\fig@scaling\linewidth]{#2.pdf} \hfil
  \includegraphics[width=\fig@scaling\linewidth]{#3.pdf} \hfil
  \includegraphics[width=\fig@scaling\linewidth]{#4.pdf}
}

\def\figtwo@scaling{0.46}

\def\showtwo#1#2{
  \centering
  \leavevmode
  \includegraphics[width=\figtwo@scaling\linewidth]{#1.pdf}
  \includegraphics[width=\figtwo@scaling\linewidth]{#2.pdf}
}

\def\figthree@scaling{0.3}
\def\showthree#1#2#3{
  \centering
  \leavevmode
  \includegraphics[height=\figthree@scaling\textwidth]{#1.pdf}
  \includegraphics[height=\figthree@scaling\textwidth]{#2.pdf}
  \includegraphics[height=\figthree@scaling\textwidth]{#3.pdf}
}

\def\figtwo@scaling{0.44}

\def\showfour#1#2#3#4{
  \centering
  \leavevmode
  \includegraphics[width=\figtwo@scaling\linewidth]{#1.pdf} \hfil
  \includegraphics[width=\figtwo@scaling\linewidth]{#2.pdf} \hfil
  \includegraphics[width=\figtwo@scaling\linewidth]{#3.pdf} \hfil
  \includegraphics[width=\figtwo@scaling\linewidth]{#4.pdf}
}

\title[Clump structure in cloud formation simulations]{Clump morphology
and evolution in MHD simulations of molecular cloud formation}

\author[R.~Banerjee et al.]{
R.~Banerjee$^1$, E.~V\'azquez-Semadeni$^2$, P.~Hennebelle$^3$ and R.S.~Klessen$^1$\\
$^1$Zentrum f\"ur Astronomie der Universit\"at Heidelberg, Institut f\"ur Theoretische Astrophysik,
  Albert-Ueberle-Str. 2, 69120 Heidelberg, Germany\\
$^2$Centro de Radioastronomia y Astrof\'\i sica (CRyA), Universidad
  Nacional Aut\'onoma de M\'exico, Morelia, Michoac\'an, Mexico\\
$^3$Laboratoire de radioastronomie millim\'etrique (UMR 8112 CNRS), \\
  \'Ecole Normale Sup\'erieure et Observatoire de Paris, 24 rue Lhomond,
  75231 Paris Cedex 05, France}

\begin{document}

\maketitle

\begin{abstract}

We study the properties of clumps formed in three-dimensional {
weakly magnetized} magneto-hydrodynamic simulations of converging flows
in the thermally bistable, warm neutral medium (WNM). We find that: (1)
Similarly to the situation in the classical two-phase medium, cold,
dense clumps form through dynamically-triggered thermal instability in
the compressed layer between the convergent flows, and are often
characterised by a sharp density jump at their boundaries { though
not always}. (2) { However, the clumps are bounded by {\it
phase-transition fronts} rather than by contact discontinuities, and
thus they grow in size and mass mainly by accretion of WNM material
through their boundaries.} (3) { The clump boundaries generally
consist of thin layers of thermally unstable gas, but these layers are
often widened by the turbulence, and penetrate deep into the clumps.}
(4) The clumps are approximately in both ram and thermal pressure
balance with their surroundings, a condition which causes their internal
Mach numbers to be comparable to the bulk Mach number of the colliding
WNM flows. (5) The clumps typically have
mean temperatures $20 \lesssim \langle T \rangle \lesssim$ 50 K, {
corresponding to the wide range of densities they contain ($ 20 \lesssim
n \lesssim 5000 \pcc$) under a nearly-isothermal equation of state}.
(6) The turbulent ram pressure fluctuations of the WNM induce density
fluctuations that then serve as seeds for local gravitational collapse
within the clumps.  (7) The velocity and magnetic fields tend to be
aligned with each other within the clumps, although both are
significantly fluctuating, suggesting that the velocity tends to stretch
and align the magnetic field with it.  (8) The typical mean field
strength in the clumps is a few times larger than that in the WNM.
(9) The magnetic field strength
in  { the densest regions within the clumps} ($n \sim 10^4 \pcc$) has a mean
value of $B 
\sim 6~\mu$G { but} with a large scatter of { nearly two orders} of
magnitude, implying that both sub- and super-critical cores are formed
in the simulation. (10) In the final stages of the
evolution the clumps' growth drives them into gravitational
instability, at which point star formation sets in, and the pressure in
the clumps' centers increases even further. 


\end{abstract}


\section{Introduction}


 The formation of molecular clouds by converging flows in the warm
neutral atomic medium (WNM) has been intensively studied in recent years
through numerical and analytical treatments \citep[e.g., ] []
{Ballesteros99b, Hennebelle99, Koyama00, Hartmann01, Koyama02, Audit05,
Heitsch05, 
Gazol05, Vazquez06, Vazquez07, Hennebelle07, Heitsch08a}, which have
shown that molecular regions can form by thermal instability TI
\citep{Field65}  in the 
{ warm neutral} interstellar medium (ISM), { nonlinearly}
triggered by transonic compressions in the WNM. { The added ram
pressure of} such compressions causes
the affected regions to overshoot from cold neutral medium (CNM) to
molecular cloud physical 
conditions. However, the nature and evolution of the clumps appearing
self-consistently within the clouds produced by this mechanism remains
controversial. On the one hand, the clumps have been reported to be in
approximate pressure balance with their surroundings and to have sharp
boundaries, as in the classical two-phase medium \citep[e.g.,
][]{Audit05, Hennebelle07}, while simultaneously, the whole medium
exhibits turbulent behavior, with wide distributions of the density and
pressure. This implies the existence of significant amounts of
gas in the unstable density and temperature ranges, which has been
suggested to be in transit between the stable warm and cold phases
\citep{Vazquez00, Gazol01, Sanchez02, Vazquez03a, Avillez04, Gazol05, Audit05}.

This coexistence of a two-phase and a turbulent regime, to which we
refer as ``thermally bistable turbulence'', is intriguing, since
turbulence implies continuous and irregular transport of gas
\citep[e.g.,][]{Klessen00b, Klessen03b}, while the two-phase regime is
usually thought to imply static conditions of the clouds, held under
confinement by the pressure of the warmer, more diffuse WNM, and with
the two phases being mediated by contact discontinuities at pressure
equilibrium, which, by definition, imply no fluid transport across
them.

%

In this contribution, we address these issues by means of
three-dimensional (3D), adaptive mesh refinement (AMR), {\it
magneto}-hydrodynamical (MHD), self-gravitating simulations, performed
with the FLASH code~\citep{FLASH00}, which allow us to have sufficient
resolution to simulate the formation of a large dense cloud complex
while still resolving the interiors of the clumps that naturally form
during this process. Note that, even though our simulations are
magnetic, in this paper we focus on the nature and evolution of the
clumps, rather than on the effect of the presence of the magnetic field,
a task that we defer to a future study.



\section{Numerical Model}
\label{sec:numerics}

We model the { convergence of WNM flows as
two colliding,} large-scale
cylindrical streams, whose evolution we follow with the FLASH code under
ideal MHD conditions. 
Our setup is similar (though not identical) to the non-magnetic
SPH simulation of 
\cite{Vazquez07} labeled L256$\Delta v0.17$ (see their Fig. 1). Each
stream is $112\, \pc$ long and 
has a radius of $32\, \pc$. They are embedded in a
$(256\,\pc)^3$ simulation box. Although the numerical box is periodic,
the cloud occupies a relatively small volume far from the
boundaries, and so the cloud can interact freely with its diffuse
environment, with relatively little effect from the boundaries.

The cylindrical streams are given an initial, slightly supersonic
inflow velocity so that they collide at the centre of the numerical
box. The inflow speed of each stream corresponds to { an isothermal}
Mach number of 
$1.22$, where the initial temperature of the atomic gas is $5000\, \K$
implying an { isothermal}\footnote{Note that
our inflow speeds are a factor of $(5/3)^{-1/2}$ smaller than 
the one in \citet{Vazquez07} as we use the isothermal sound speed,
while those authors used the adiabatic one.} sound speed of $5.7\, \km\,
\sec^{-1}$.  We also add 10\% random velocity 
perturbations to the bulk stream speeds. The initially homogeneous
density is $n = 1 \, \cm^{-3}$ ($\rho = 2.12\times 10^{-24} \, \g~
\cm^{-3}$, { using a mean atomic weight of 1.27}). 

Here, we report on the results from our weakly magnetised case with a
homogeneous magnetic field component of strength $B_x = 1\, \mu\G$,
which is parallel to the gas streams. These initial conditions
translate to a plasma $\beta = P_{\sm{therm}}/P_{\sm{mag}}$ of
$17.34$. We choose this relatively weak field so that the gas in the
streams is magnetically supercritical, and thus the cloud formed by
compression along the field lines can eventually also become
supercritical. Note that the only way in which our clouds can become
supercritical is by accreting mass from the inflows along the field
lines \citep{Mestel85, Hennebelle00, Hartmann01, Shu07, Vazquez07b},
since we do not include ambipolar diffusion in our simulations. In any
case our choice of the initial uniform field is not overly low, as
recent estimates of the mean Galactic field give upper and lower
limits of $4\pm 1$ and $1.4 \pm 0.2~\mu$G, respectively
\citep{Beck01}. On the other hand, total magnetic field strengths
which also comprise of field fluctuations are typically reported to be
of the order of $6\, \mu\G$ \citep[see e.g.,][]{Heiles05}. Here, we
are not including initial field fluctuations. In our simulation, such
fluctuations are dynamically generated in the CNM from the initial
turbulent velocity and thermal instability.


This setup is different from the numerical setup in our companion paper,
\citet{Hennebelle08b}, where a significantly stronger field was used ($5
\mu$G), and the colliding streams entered the simulation box along the
field direction through two opposite boundaries, which had inflow rather
than periodic conditions. Thus, in that paper, the mass-to-flux ratio of
the cloud was allowed to increase without limit, while in our
simulations, this ratio is bounded from above by the mass-to-flux ratio
of the simulation box.  

These initial conditions result in a ratio of the total mass in the
flows, $M_{\sm{flows}} \approx 2.26\times 10^4 \, \Msol$, to the
thermal Jeans mass of $M_{\sm{flows}}/M_J \approx 2\times
10^{-3}$. The warm neutral medium is far from being gravitationally
unstable. On the other hand, the mass-to-flux ratio of the flows, $\mu =
M_{\sm{flows}}/\Phi_B = \Sigma/B_x$, is such that $\mu/\mu_{\sm{crit}}
\approx 2.9$, where $\mu_{\sm{crit}} \approx 0.13/\sqrt{G}$ is the
critical value for a slightly flattened structure on the verge of
collapse~\citep{Mouschovias76, SpitzerBook78}. Using this critical value, the mass-to-flux ratio of the {\em entire} simulation box is $3.3\, \mu_{\sm{crit}}$, because the extend of the flows is slightly smaller than the box size.
We estimate the
transition time at which the cloud becomes magnetically super-critical
as
\bea
t_{\sm{trans}} & \approx & 5.4\, \Myr \nonumber \\ 
  & & \times \left(\frac{n}{1\,\cm^{-3}}\right)^{-1}
  \left(\frac{v_{\sm{flow}}}{6.9\, \km \, \sec^{-1}}\right)^{-1}
  \left(\frac{B_x}{1\, \mu\G}\right).
\label{eq:crit_time}
\eea
{ Concerning the cooling}, we use the procedure { described} in
\citet{Vazquez07}, which is based on the chemistry and cooling
calculations of \citet{Koyama00} and the analytic fits to them by
\citet{Koyama02}. { Thermal conduction is neglected.}

We follow the gravitational collapse with up to 11 AMR refinement
levels, which correspond to a { maximum} resolution of 8192 grid
points~\footnote{Note, each refinement level corresponds to a block $8^3$ grid points}, or a grid spacing of $\Delta x = 0.03\,\pc$ in each
direction. For the dynamical mesh refinement we use a Jeans'
criterion, where we resolve the local Jeans' length with at
least 10 grid cells~\citep[see][for the necessary criterion to prevent
artificial fragmentation]{Truelove97}.  { It should be noted that
some authors have advocated that the resolution criterion should be to
resolve the Field length \citep{Field65} with at least three grid points
\citep{Koyama04, Gressel09}. Other authors \citep{Hennebelle07c} have suggested that the
proper scale that needs to be resolved is the ``sound crossing scale'',
the product of the sound speed and the cooling time. In the warm and
cold atomic phases, either of these criteria may actually be more
stringent than the Jeans one but, as also discussed by \citet{Hennebelle07c}, the
main effect of insufficient resolution in this kind of simulations is
to truncate the fragmentation at the smallest resolved scales, in turn causing the the low-mass end of the clump mass spectrum to be also truncated. 
However, in this paper we are not concerned with resolving small scale
fluctuations nor in the detailed shape and extension of the core
spectrum, but rather with the mechanism of clump growth, and the
clumps' substructure, once they have reached well-resolved sizes.
As discussed by \citet{Vazquez06}, the physics of clump
growth are simple, and do not require a very high resolution. In that
paper, it was also concluded that, since clumps grow, they
are necessarily unresolved during the initial stages of their formation,
but later become well resolved, as they reach sufficiently large
sizes. Thus, we can investigate the substructure of clumps that have had
enough time to become well resolved. 
Nevertheless, it could be that by also resolving small scale fluctuations the conversion of the WNM to the CNM might be slightly faster if the surface-to-volume ratio of the cold gas increases in this case.
On the other hand, since we plan
to use the present simulations to study the star formation process in
future contributions, it is important to satisfy the Jeans criterion. In
any case, our clumps are typically resolved with at least the 10th level
of refinement, corresponding to a resolution of 0.06 pc.}

Finally, Lagrangian sink particles \citep[see e.g.,][]{Bate95,
  Jappsen05} are created if the local density exceeds $n > n_{\rm sink}
\equiv 2\times 10^4 \pcc$ and this location is a local minimum of the
gravitational potential. These particles interact only gravitationally
with the gas and with each other. They are free to move within the
simulation box independent of the underlying mesh (i.e Lagrangian
particles). The sink particles have an accretion radius of $0.47\,
\pc$ corresponding to roughly one Jeans' length at $n_{\rm
  sink}$. Within this accretion radius, gas in excess of $n_{\rm
  sink}$ is deposited into the sink if this gas is gravitationally
bound. The sink mass increases acordingly. Note that the initial mass
of the sink is computed with the dynamically accreted mass; i.e., only
the mass in excess of $n_{\rm sink}$ contributes to the initial sink
mass.

In a forthcoming study we investigate the influence of ambipolar
diffusion (AD) on molecular cloud formation in colliding flows, where
we are using the implementation of ambipolar diffusion for the FLASH
code by \citet{Duffin08}. Ambipolar diffusion might regulate the
magnetic field strength in the condensations that are induced by
thermal instability \citep{Inoue07, Inoue08}.


\section{Results}
\label{sec:results}

\subsection{Global features} \label{sec:global}

\befone
\showtwo{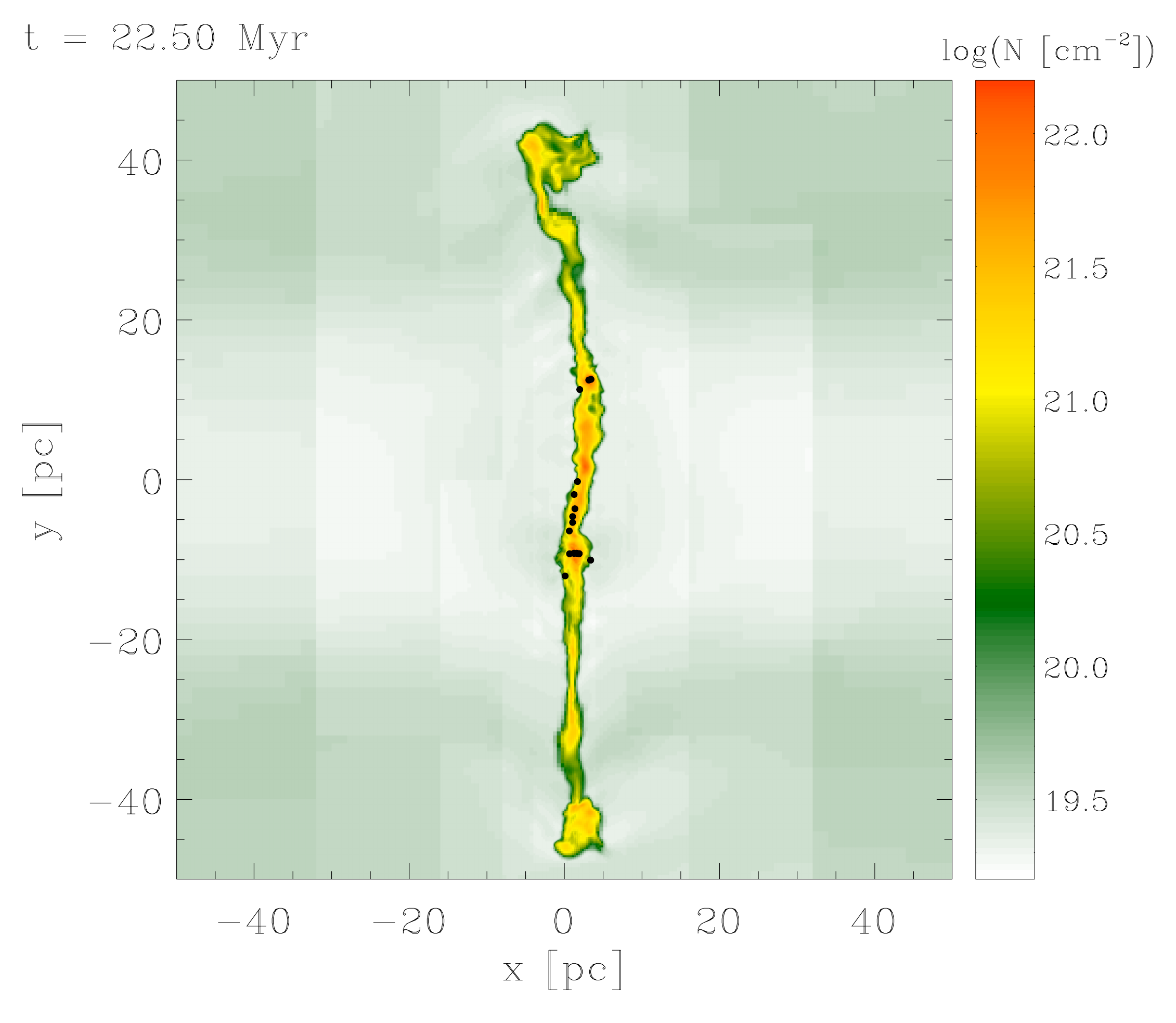}
	{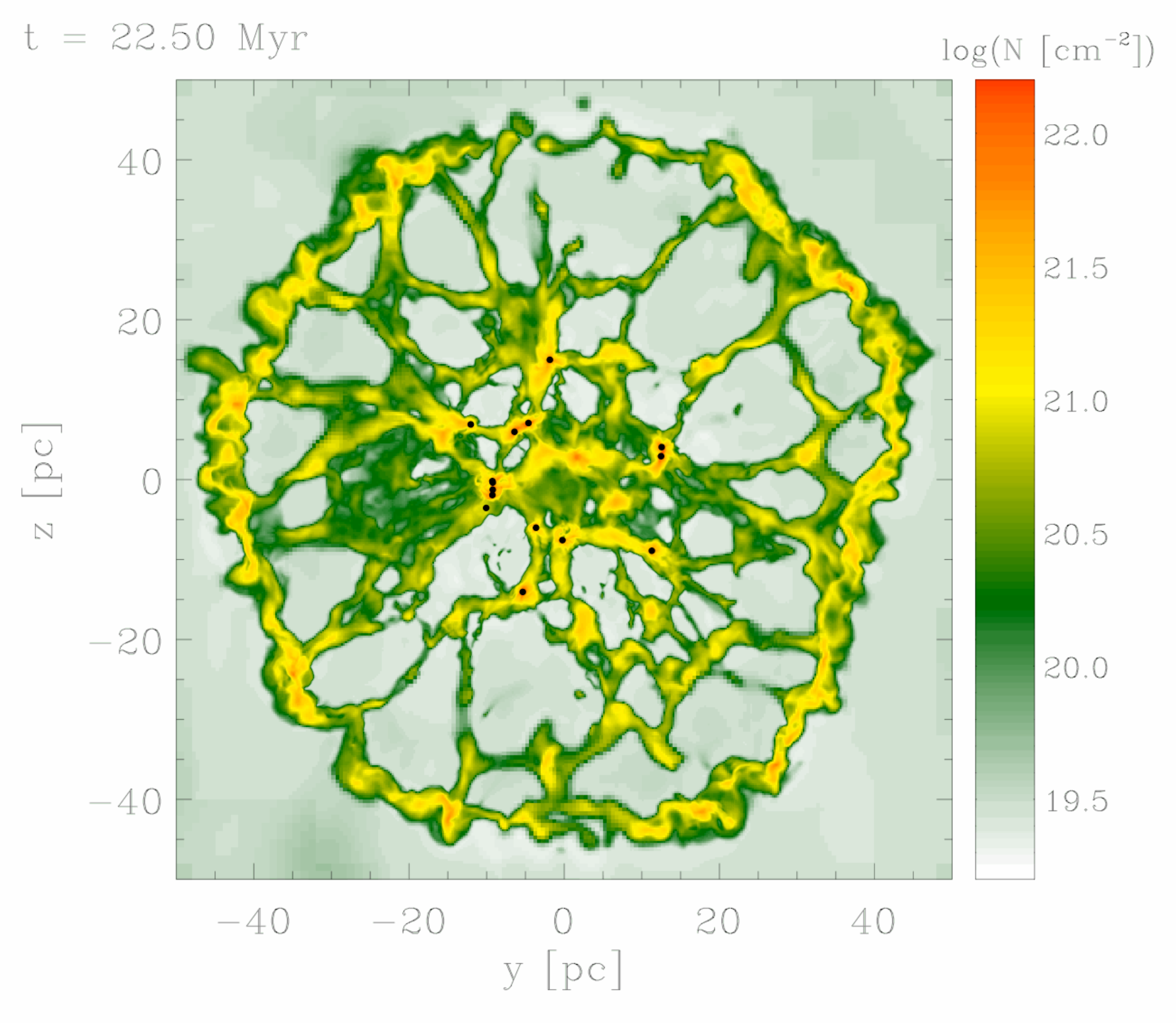}
\caption{Column density of the inner region of the molecular
  cloud viewed edge-on (left panel) and face-on (right panel). In
  the left panel, the large scale flows advance inwards from the
  right and left sides of the box, leaving a lower-density medium
  there. In both panels, the dots mark the projected positions of
  the sink particles, i.e. regions of local gravitational
  collapse. The first regions that show active star formation
  appear { at} about $17\, \Myr$ in this simulation. The
  molecular cloud is largely inhomogeneous, with the ``molecular''
  gas ($\log(N/\cm^{-2}) \simge 20.5$) interspersed within the
  warm atomic gas. Note that the simulation box is 4 times larger
  (i.e. $256 \, \pc$ each side) than the area shown here.}
\label{fig:cloud}
\eefone

As is already well known, the collision of transonic,
converging flows initially produces moderate compressions on the
{\it linearly stable} WNM at the collision front, which are
sufficiently strong to  nonlinearly trigger thermal instability
\citep{Hennebelle99}, so that the gas rapidly cools to temperatures well below 
$100\, \K$, forming a thin sheet that then fragments into filaments and
ultimately into small clumps. Moreover, the
thermal pressure  of the dense gas is in close balance with the
thermal + ram 
pressure of the WNM outside { it}, and { it is} at higher
densities and pressures than the { steady-state CNM,}
overshooting to { typical giant} molecular cloud { physical
(density and temperature)} conditions \citep[$n \sim 100 \pcc$, $T \sim
$ a few tens of Kelvins]{Vazquez06}. This process
also causes the newly formed 
dense gas to be turbulent, with a transonic velocity dispersion with
respect to its own sound speed \citep{Koyama02, Heitsch05, Vazquez06,
Hennebelle08b}.  As in the classical two-phase model
\citep{Field69}, the clumps { in the dense gas} are bounded by sharp
density jumps of roughly a factor of 100. We refer to
this regime of global turbulence subjected to the tendency to form dense
clumps by TI as ``thermally bistable turbulence''.  { In what
follows, we}
indistinctly refer to the dense { clumps} as ``molecular'', even
though we do not actually follow the chemistry in our simulations.

In this regime, the ``molecular cloud'' is
composed of a mixture of diffuse and dense gas,
with a significant fraction of gas in the unstable range, which is in
transit from the diffuse to the dense phase, thus
producing a continuous though bimodal distribution of densities and
temperatures \citep{Vazquez00, Gazol01, Gazol05, Audit05}. In
Fig.~\ref{fig:cloud} we show the structure of the 
{ dense} cloud from our converging flow simulation about $5\, \Myr$
after the first regions collapse and form stars. In particular, from
the face-on image one can see that the ``cloud'' is not a 
homogeneous entity, but rather it is composed of dense clumps (which
should be mostly molecular) embedded in somewhat less dense filaments
(which may be partly molecular and partly atomic) \citep[see
also][]{Vazquez07, Hennebelle08b}.

\befone 
\showthree{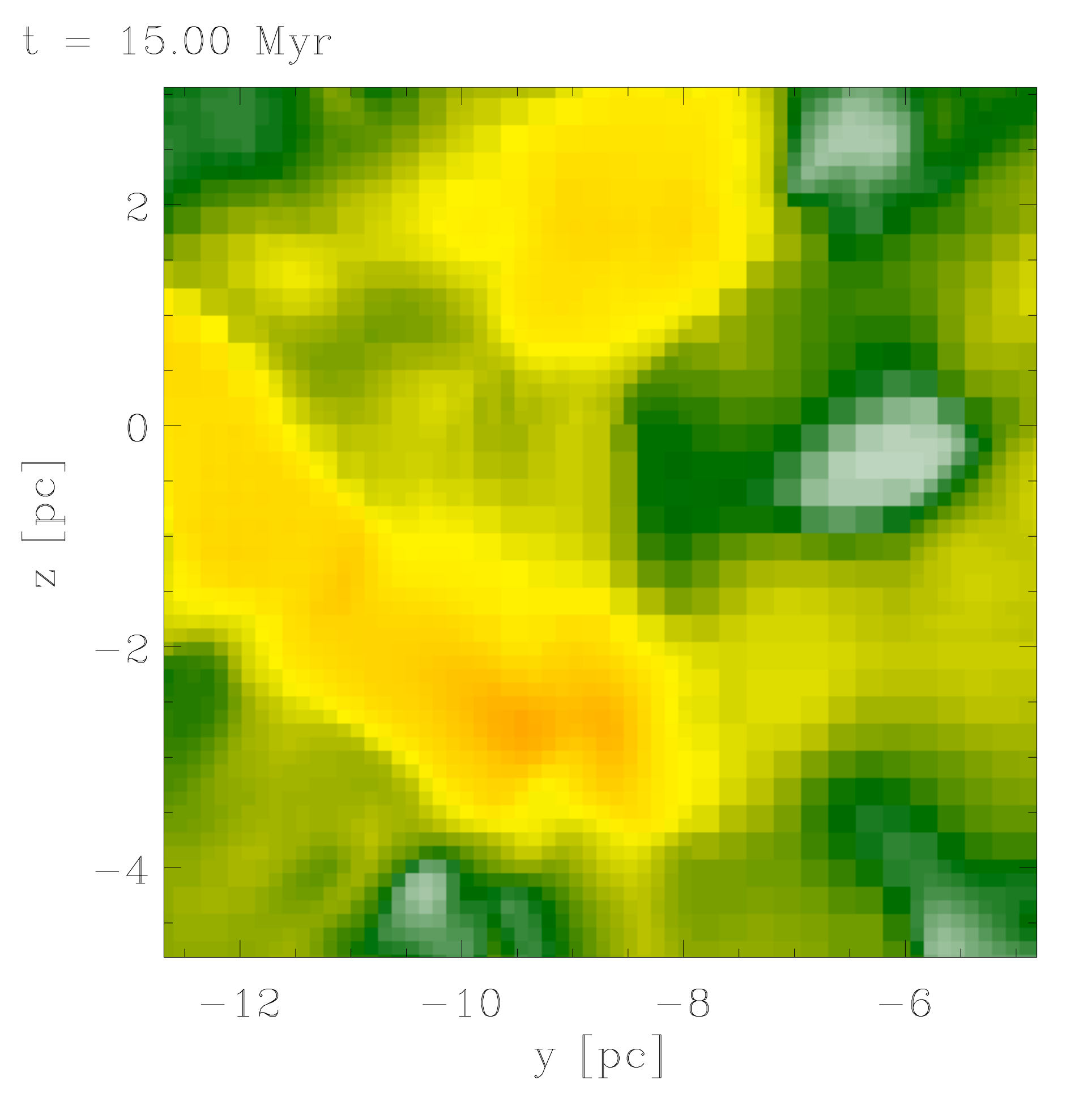}
          {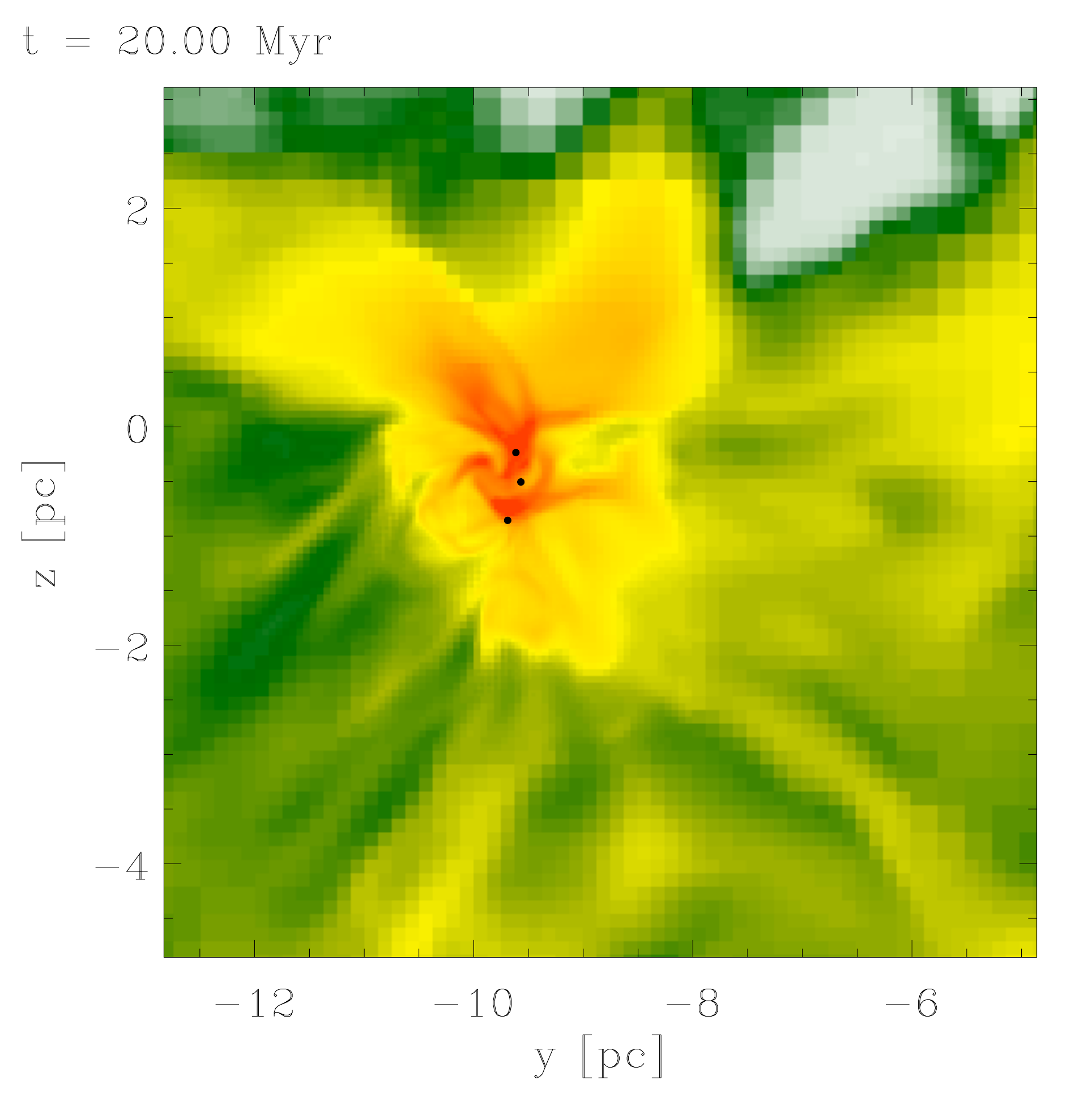}
          {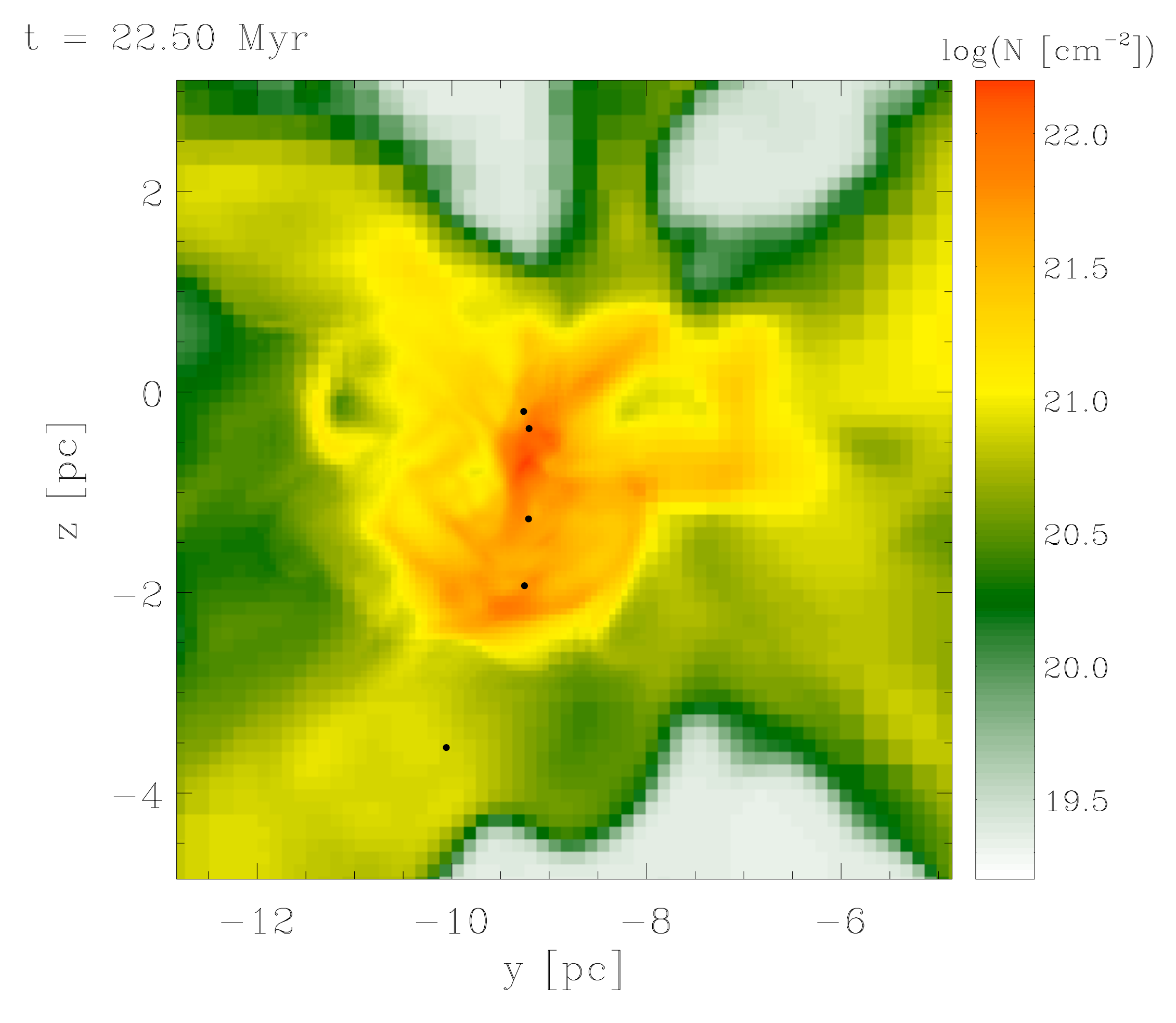}
\caption{Column density evolution of a  clump (colored
yellow-green) in which
a self gravitating, cold, dense core (yellow-red) builds up. 
The clump is embedded in a large-scale, filamentary structure (see also
Fig.~\ref{fig:cloud}) but  is separated from the external WNM
(light-blue) by a
sharp boundary (dark green), although it exhibits a complex, fluctuating
substructure. The clump grows mainly by accretion of material 
from the WNM. Cold, dense regions become Jeans unstable and start to
form stars (indicated by the black dots). See also
Fig.~\ref{fig:clump_structure_center}.}
\label{fig:clump_evol}
\eefone

\bef
\showone{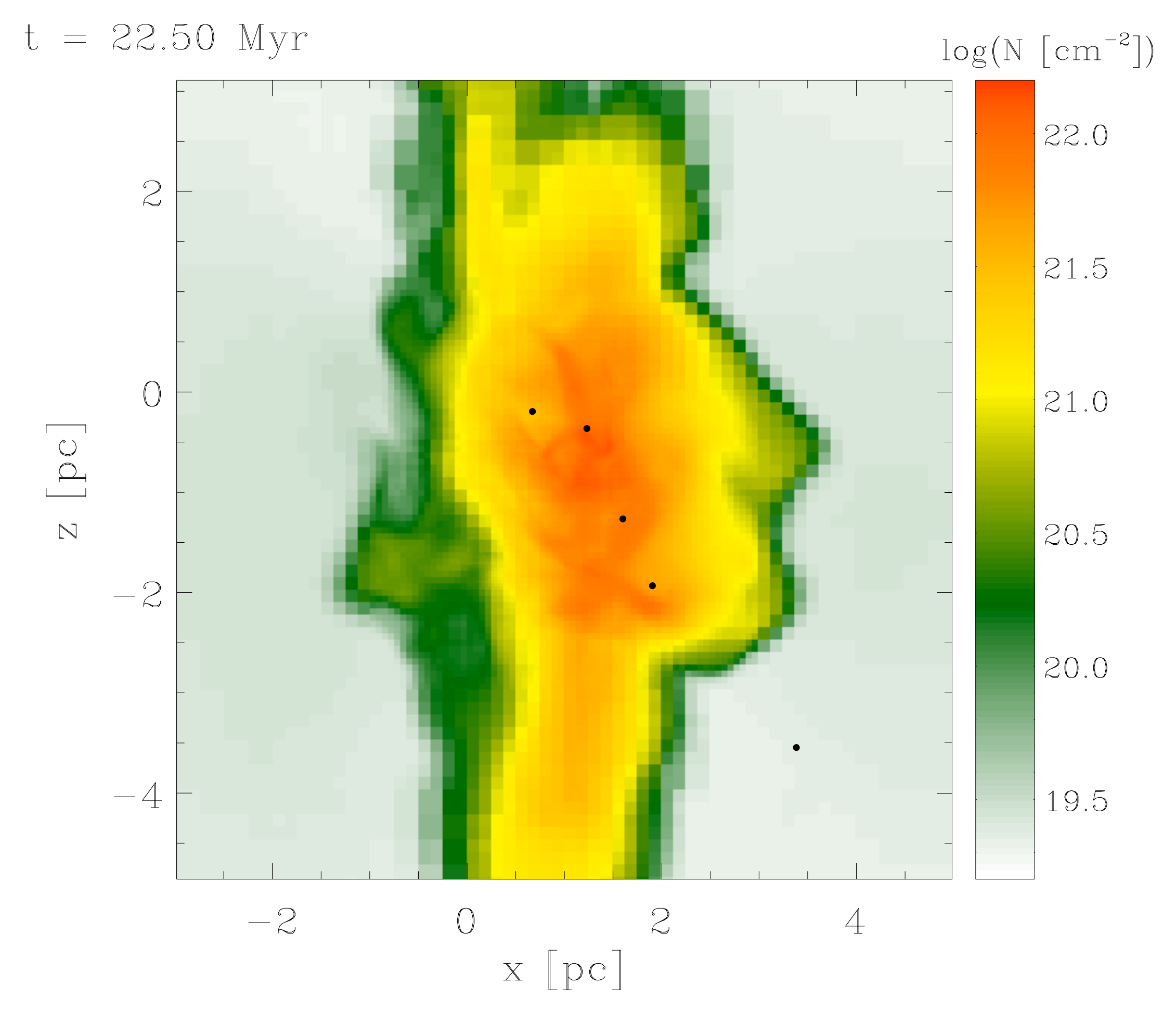}
\caption{Column density of the edge-on view of the clump shown in
  Fig.~\ref{fig:clump_evol}. The clump clearly originates at the
  intersection of the large scale flows and is connected to a large
  filamentary structure. Perpendicualar to the filament the clump
  grows through the propagation of its phase transition front into the
  WNM. }
\label{fig:clump2_edge}
\eef

\bef
\showone{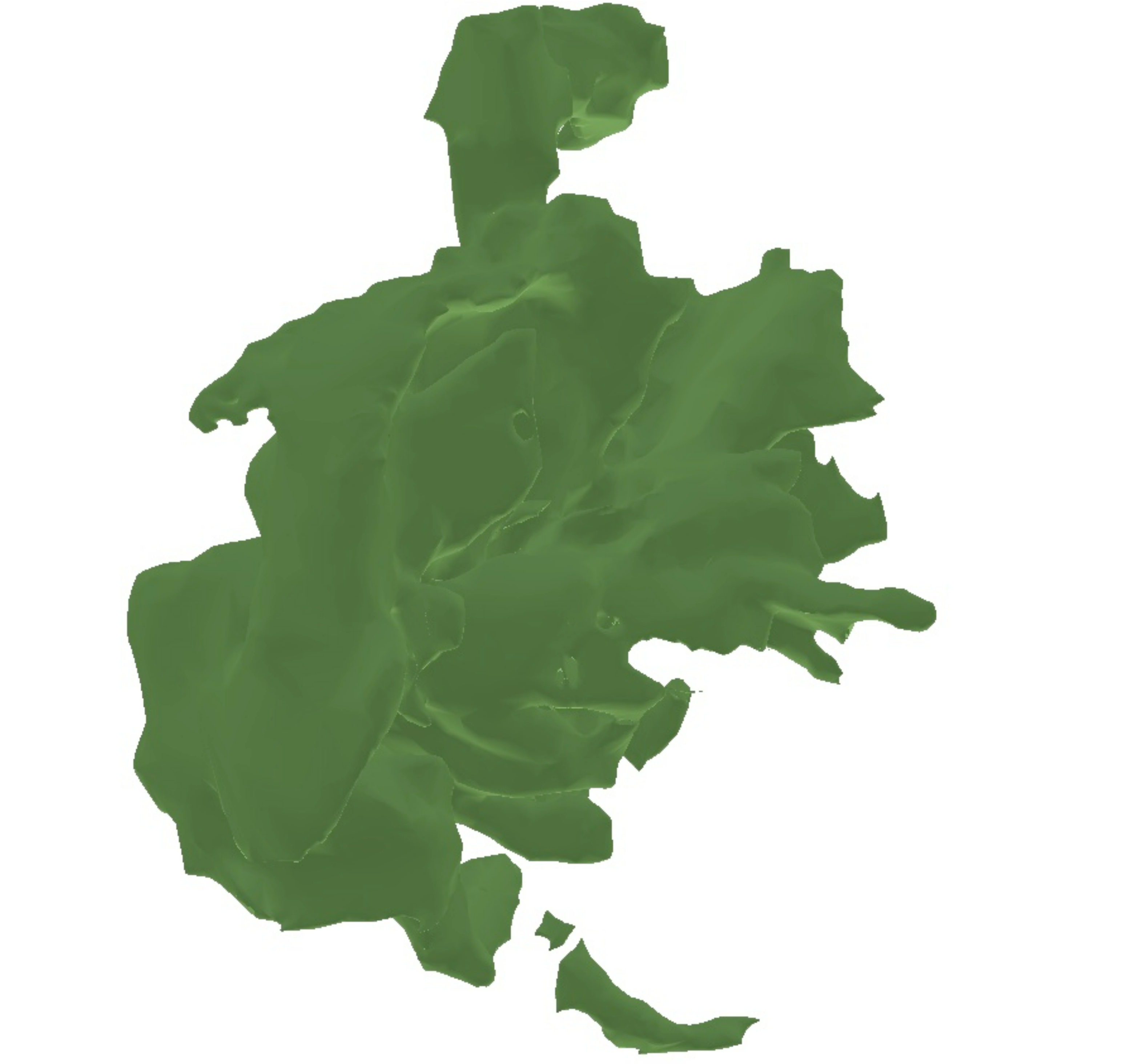}
\caption{Three dimensional structure of the clump shown in the last
  panel of Fig.~\ref{fig:clump_evol}. The density iso-surface is shown
  for number density of $600\, \cm^{-3}$ reveals the complex structure
  of this clump.}
\label{fig:clump2_3D}
\eef

\subsection{Clump properties} \label{sec:clumps}

\subsubsection{Clump growth mechanism}

As mentioned above, the clumps are mainly bounded by sharp density
jumps of roughly a factor of 100. However, the clumps are connected to
large filaments which build up in the intersection region of the
colliding flows. The main difference between the regime in the
simulations and the classical two-phase model of \citet{Field69} is
that, because the clumps are formed by turbulent compressions in the
WNM rather than by linear development of TI, {\it they can accrete mass
from distances much larger than the scale of the fastest-growing mode
of TI in the diffuse medium}, which is typically small. In a turbulent
environment, instead, the clumps can accrete from the scales
associated with the compressive motion that forms the clumps. 
This implies that, for turbulence-induced clump formation, the duration
of clump growth may be much longer than in the case of linear
development, and thus one should expect to generally find the clumps in
a growing stage, rather than in a { quasi-static state, as in the
final state of the linear development of TI}. In turn, this
means that there should { generally} be a net mass flux across their
boundaries {
driven by the {\em ram} pressure of the inflow.
This dynamic growth is different to the situation in the quasi-static two-phase model, in
which any mass flux through the fronts \citep[evaporation or
condensation; e.g.,][]{Zeldovich69} occurs as a result of the tendency to equalize
the {\it thermal} pressure between the CNM and the WNM. We thus refer to
the clumps' boundaries as  {\it phase transition fronts}.}

This mechanism has
been studied in detail in one dimension by \citet{Hennebelle99} and
\citet{Vazquez06}. The latter authors gave analytical
solutions for the expansion velocity of the phase transition front that
separates the warm and cold gas as a function of the bulk Mach number of
the inflows (see their fig. 3). Typically, those speeds are small,
because the gas is tightly packaged in the
clumps, and so the clump size increases slowly. In fact,
interestingly, the front speed {\it decreases} with increasing inflow
Mach number because the ram pressure of the compression then causes the
clump density to be { sufficiently large as to overwhelm the larger
accretion rate onto the cloud}, and the net effect is that the {
front propagates at} lower speeds. 
During their growth, the clumps also ocassionally coalesce with
other nearby clumps, a process that enhances their growth rate. 

In Fig.~\ref{fig:clump_evol} we show the { column density} evolution
of a typical clump
in our simulation, which exhibits the aforementioned sharp boundaries
and growth, { although large fluctuations are seen within it as
well}. A few million years later, the clump becomes 
self-gravitating and starts to { produce local sites of} collapse. We
also { show an edge-on column density view} of the clump at $t =
22.5\, \Myr$ in 
Fig. \ref{fig:clump2_edge} and { a} 3D iso-density image in
Fig.~\ref{fig:clump2_3D}, which shows the complexity of the dense
clump. 

It is important to note that { the clumps' growth} mechanism is a simple
consequence of the thermal bistability of the flow, triggered by the
transonic compression in the WNM, so it is essentially a
thermo-hydrodynamic phenomenon, { independent of the presence of the
magnetic field}. Indeed, Fig.~\ref{fig:clump_hydro}
shows the density and velocity field for a clump in a non-magnetic
simulation, showing that the same growth mechanism occurs there as
well. The inclusion of the magnetic field does not appear to change
its basic action, as the gas simply tends to flow along field lines,
which in turn are re-oriented by the inertia of the flow
(cf. sec. \ref{sec:int_struc}).

\bef
\showone{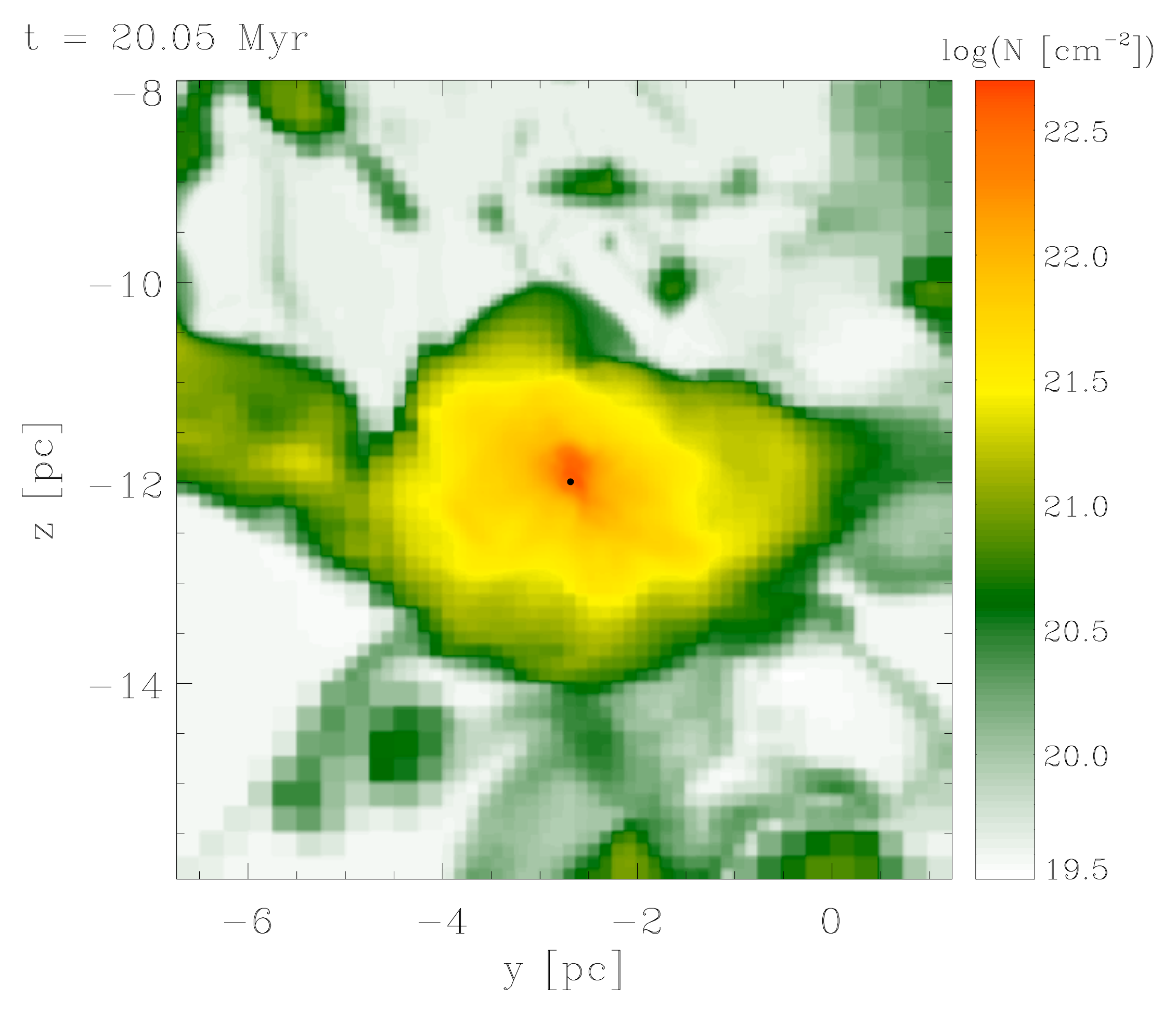}
\caption{Column density of a clump from
  the hydro simulation. The structure of the clump { and its
growth by accretion of WNM material are}  very
  similar to the weakly magnetised case { shown} in 
  Fig.~\ref{fig:clump_evol}.}
\label{fig:clump_hydro}
\eef

\befone
\showfour{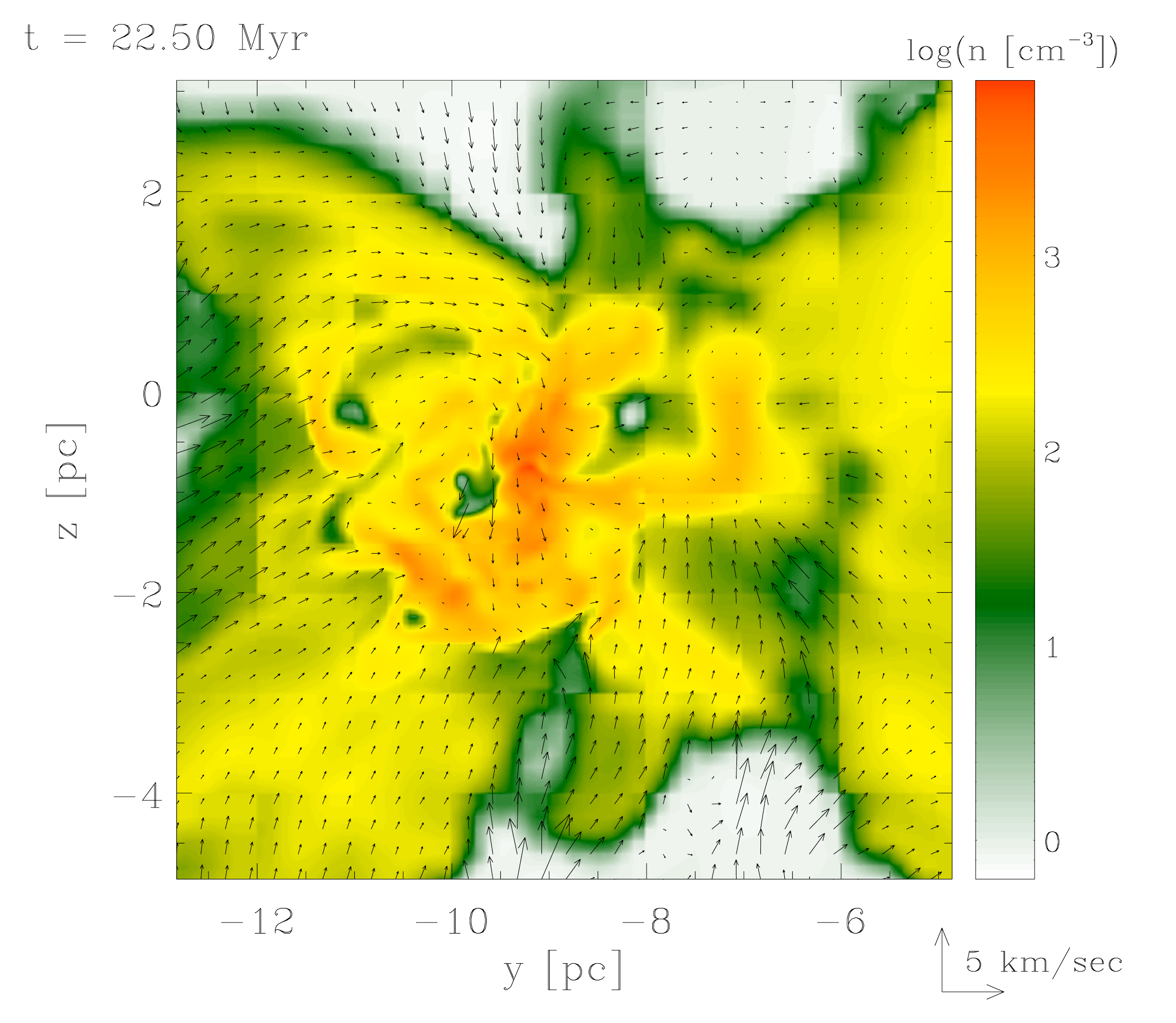}
          {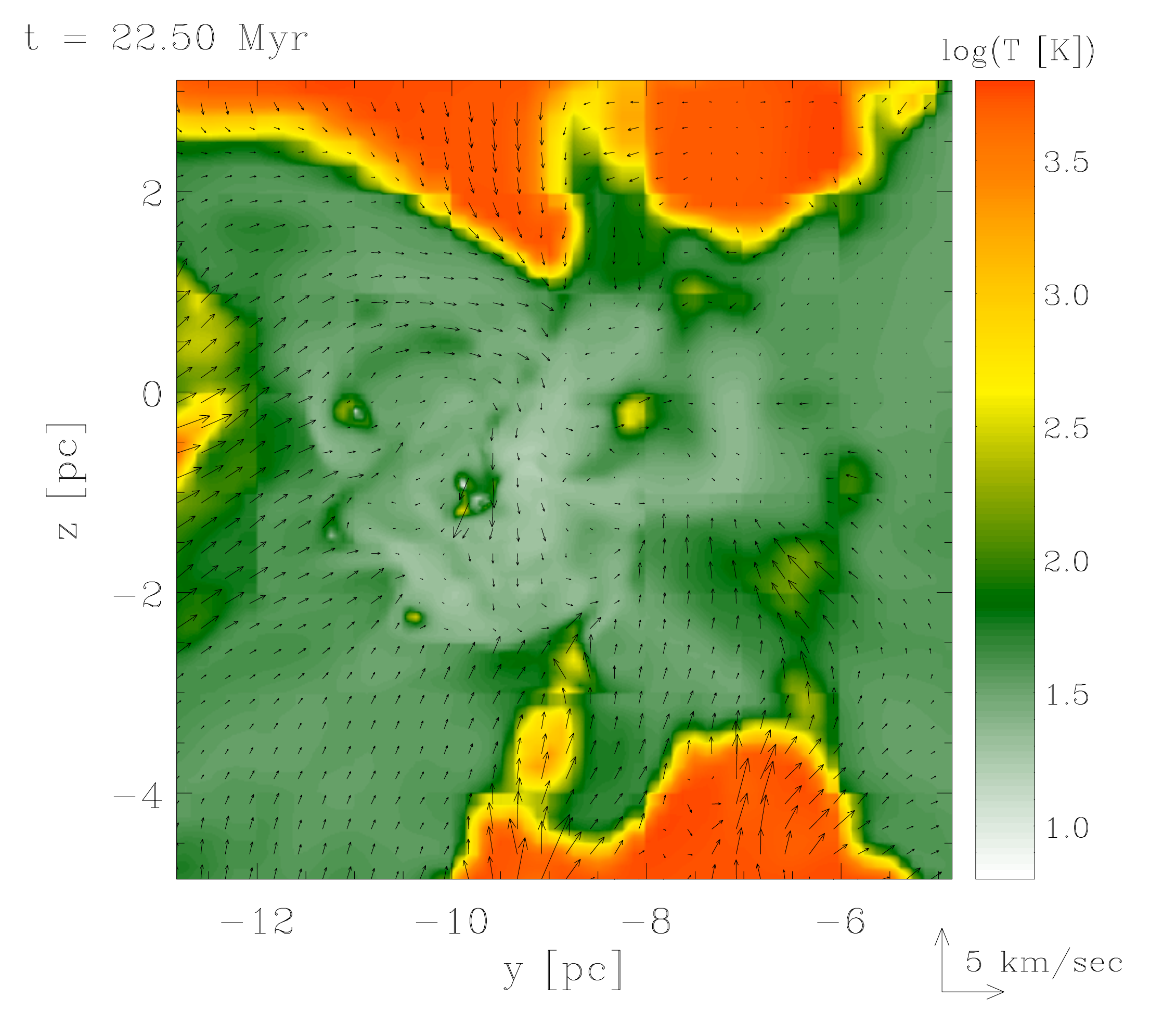}
          {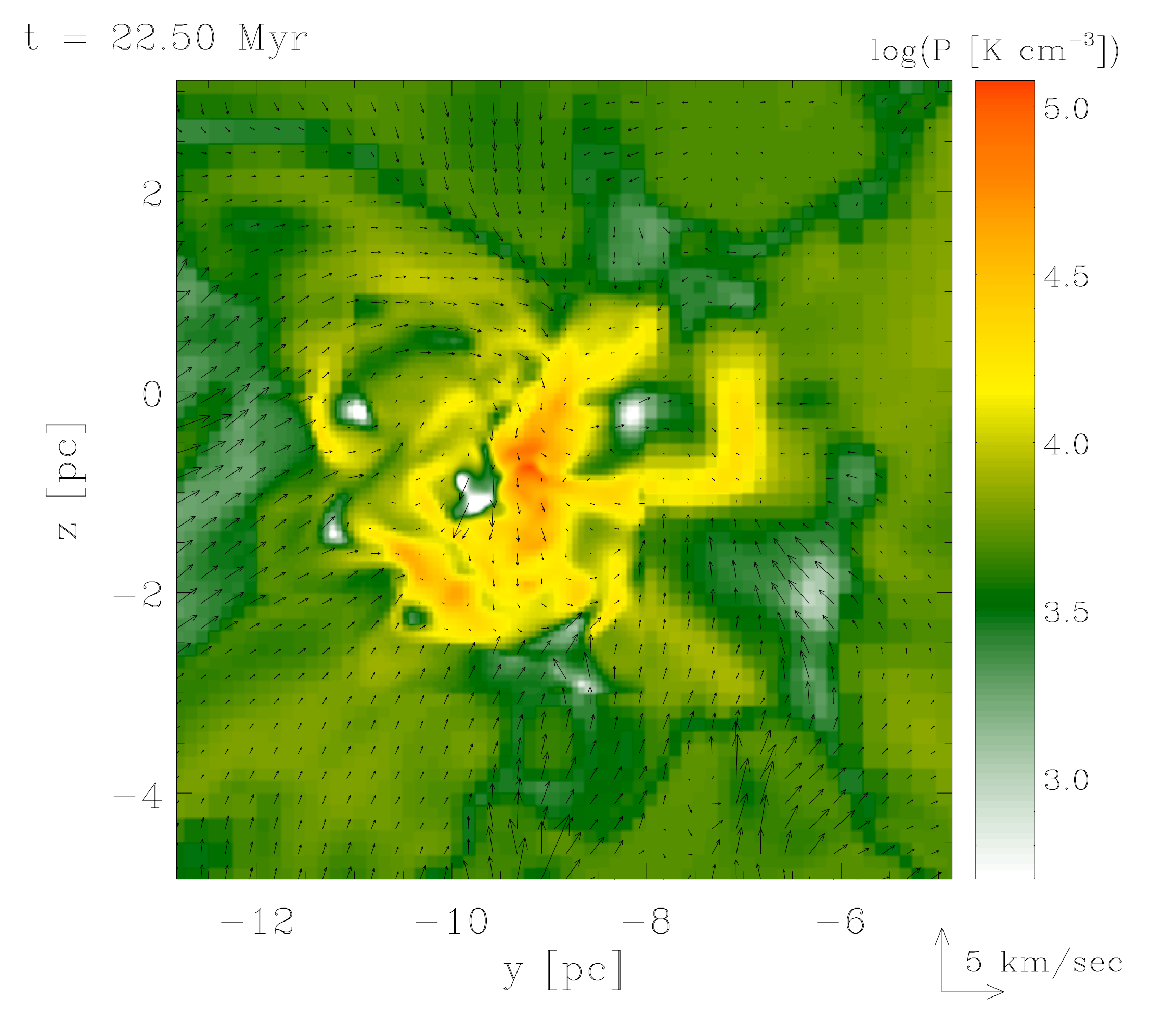}
          {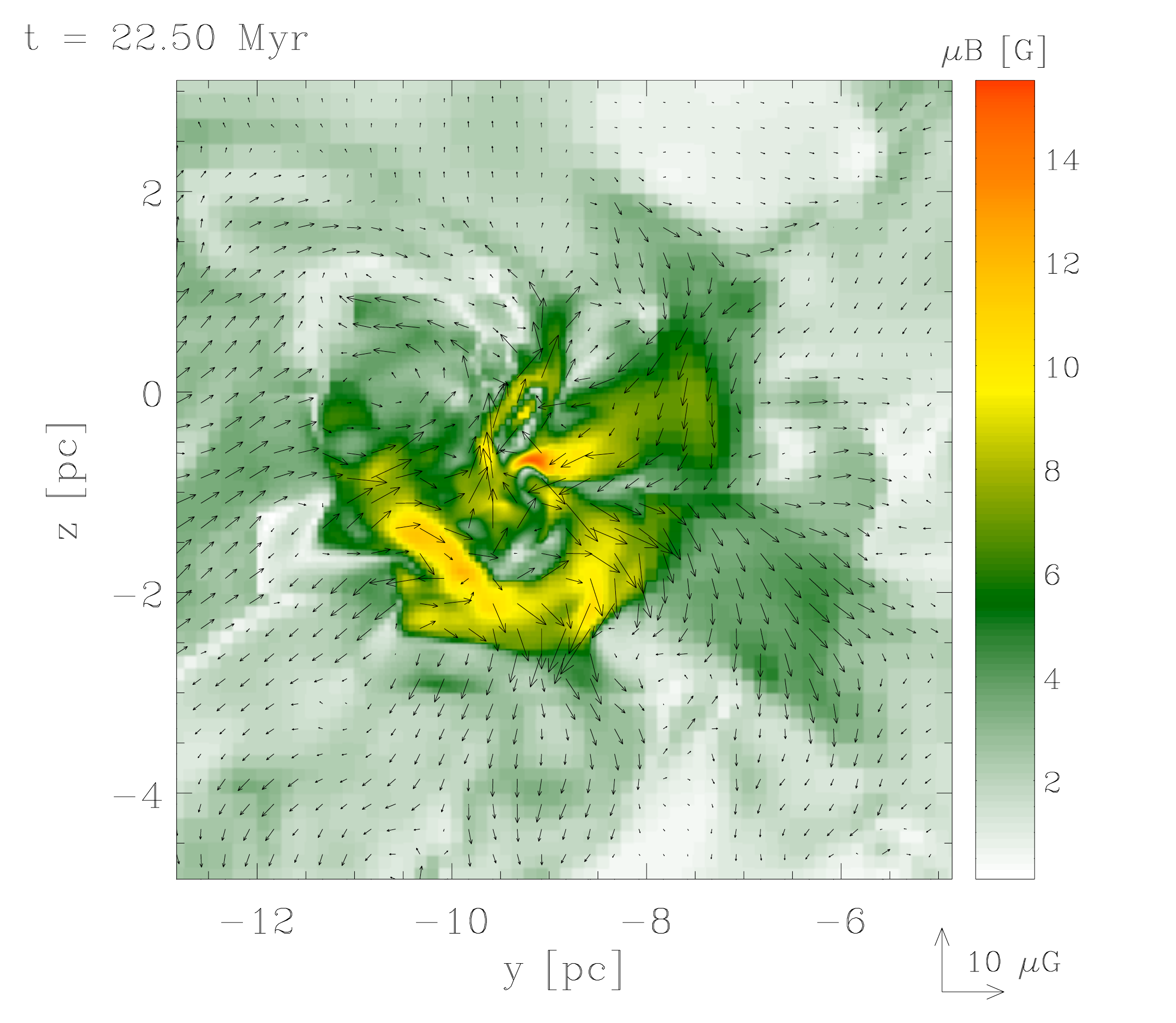} \caption{{ Structure
          of the clump shown in the right panel of
          Fig. \ref{fig:clump_evol}}. The images show 2D slices through
          the densest region of the clump. Shown are the { density
          (top left), temperature (top right), thermal pressure (bottom
          left), and magnetic field strength (bottom right; note the
          linear scale)}. The arrows in the pressure image indicate the
          velocity field and, in the magnetic field image, the magnetic
          { field vectors}.
            The 
WNM gas streams into the clump predominately along the magnetic flux
lines. { Note in the top left panel that the clump boundaries (dark
green) are generally thin, but on occasions become wide and penetrate
deep into the core, causing the transition from WNM to CNM to 
``molecular'' gas to be smoother.}}
\label{fig:clump_structure_center}
\eefone

\befone
\showfour{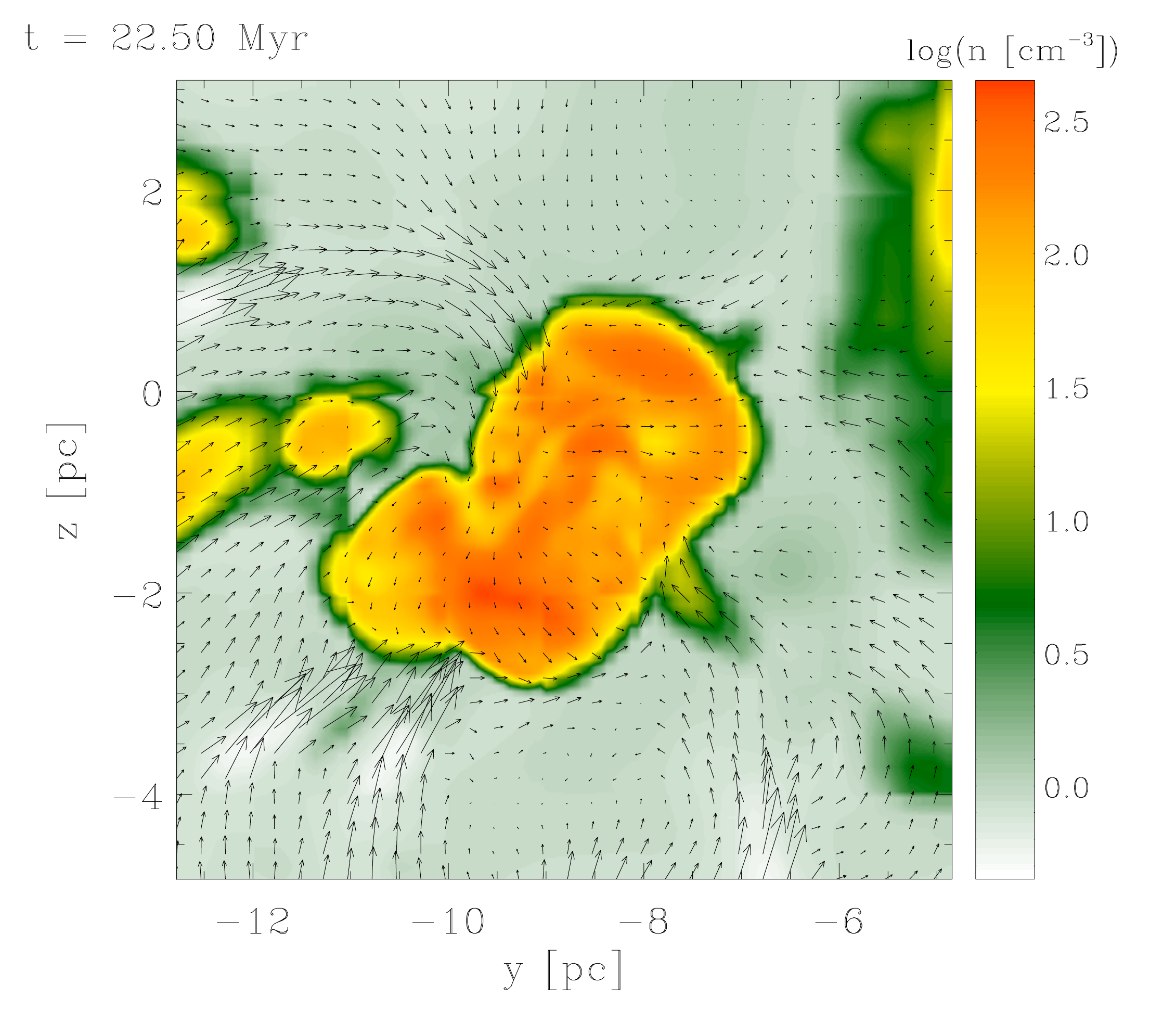}
          {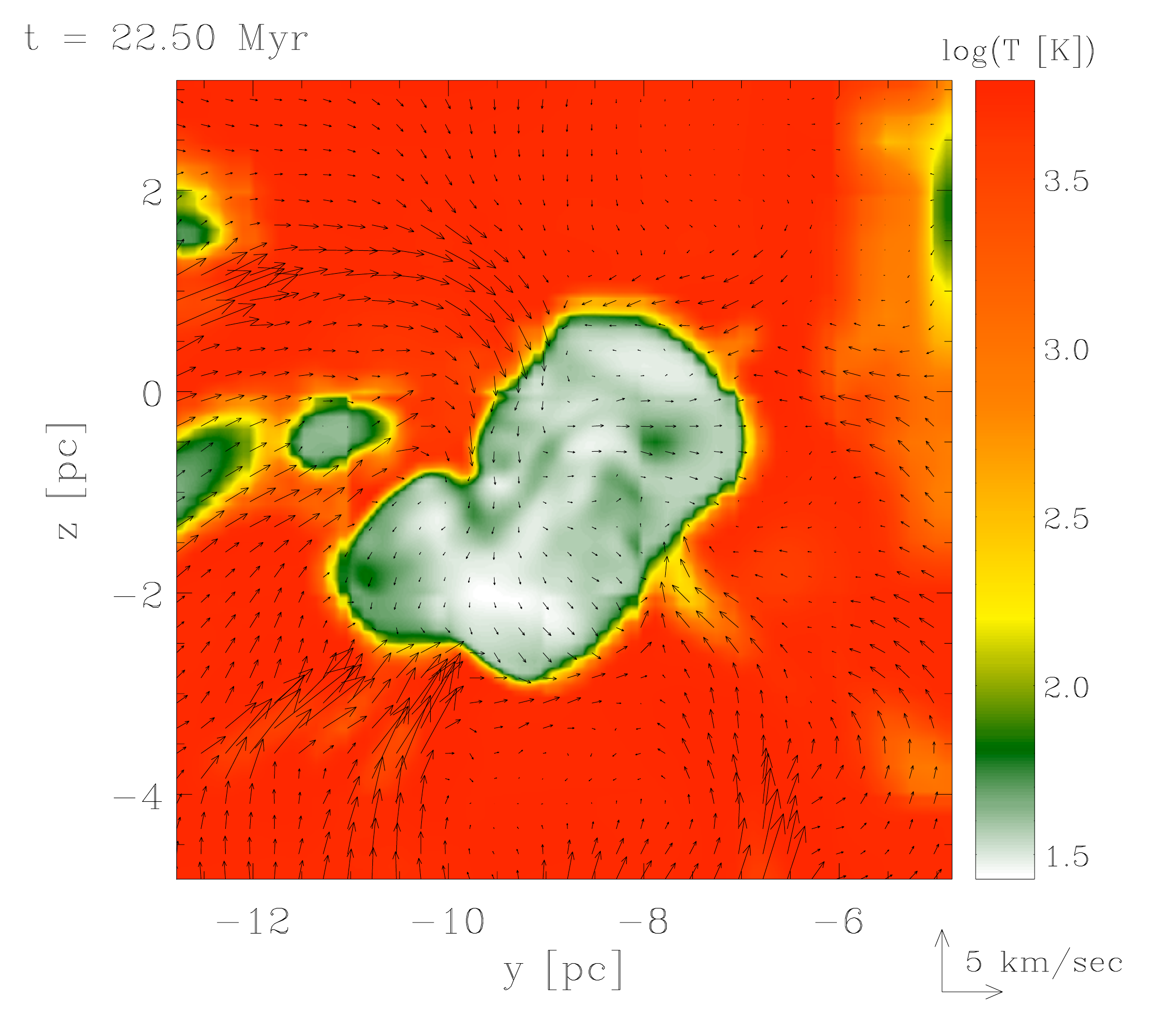}
          {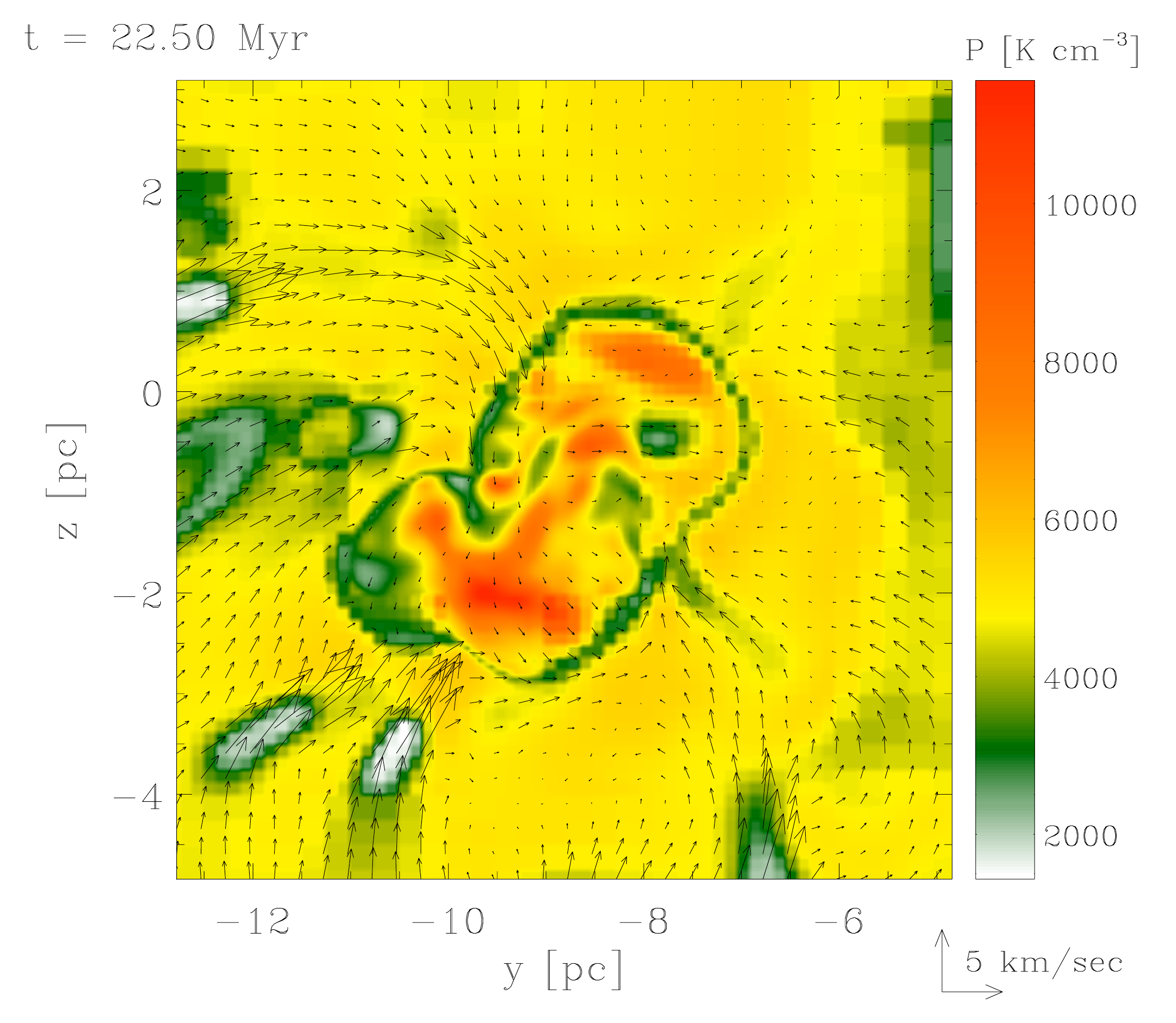}
          {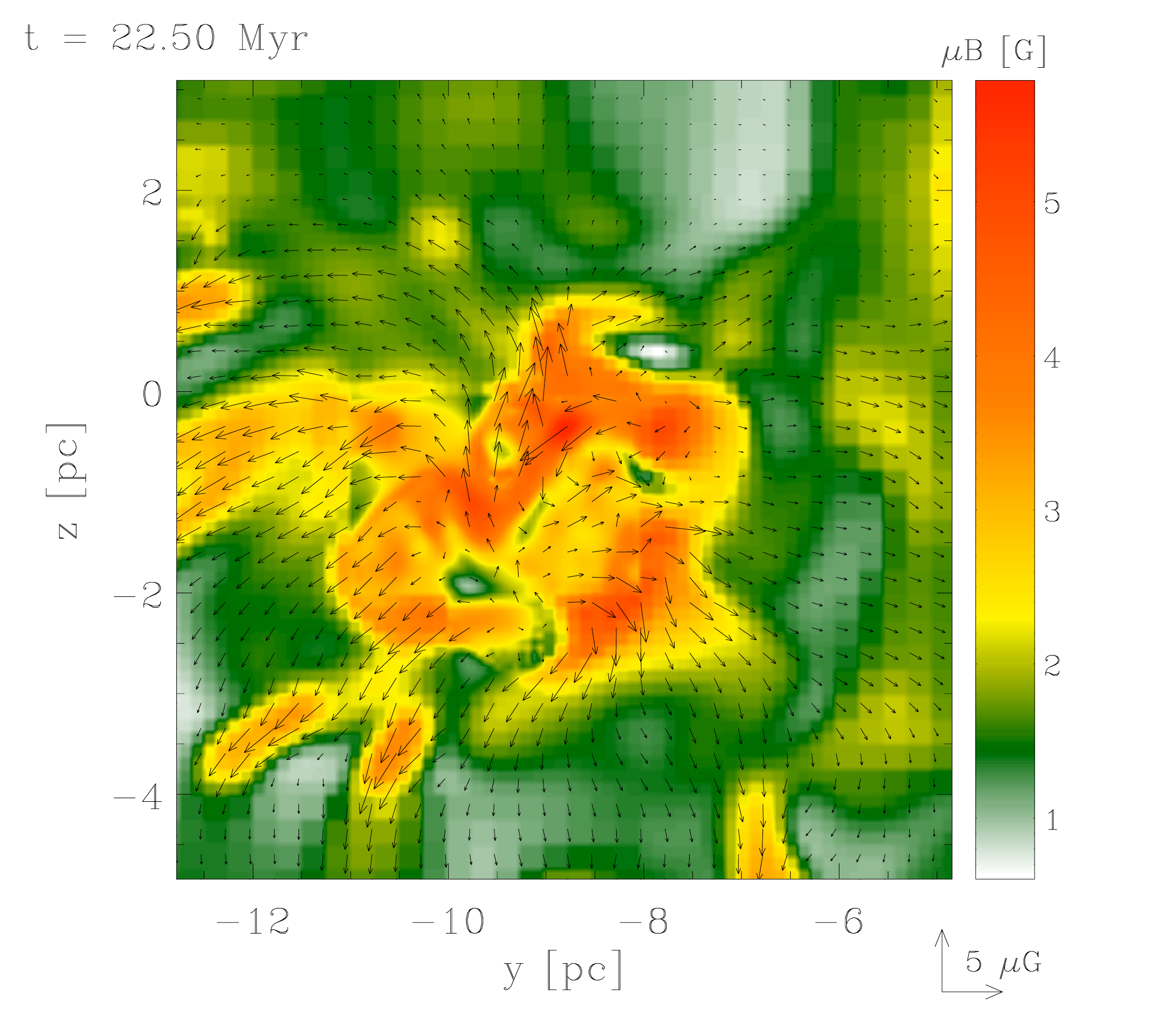}	 
          \caption{Same 2D images than shown in
            Fig.~\ref{fig:clump_structure_center} but cut through 
            slightly off center ($x = 2.5 \, \pc$, see also
            Fig.~\ref{fig:clump2_edge}). Here the clump properties are
            highly distinct. In particular, the density and
            temperature contrast compared to the WNM is greatly
            prominent. Due to the thermal { bistability of the
              flow, the clump is} almost in pressure equilibrium with
            { its} surroundings.}
\label{fig:clump_structure}
\eefone

\subsubsection{Clump internal structure} \label{sec:int_struc}

To examine the internal structure of the dense clumps, in
Figs.~\ref{fig:clump_structure_center} and \ref{fig:clump_structure} we
show { slices through the clump shown in the last panel of
Fig. \ref{fig:clump_evol}, at two different $x$ positions, one
through the site of its maximum density (Fig.\
\ref{fig:clump_structure_center}), and the other closer to its periphery
(Fig.\ \ref{fig:clump_structure}). The four
panels of each figure respectively show maps of the density},
temperature, pressure, and magnetic field strength of the clump. The
resolution in the interior of the clump is 0.03 pc, while its linear
dimensions at $t=22.5$ Myr are seen to be roughly $3 \times 5$ pc, {
suggesting that the clump is well resolved}. 

{ At this point, it is convenient to estimate the expected
pressure-equilibrium value of the density in the clumps, when one
includes the ram pressure from the colliding inflows. The density and
temperature initial conditions of our simulation imply a thermal
pressure $P_{\rm th} = 5000~\K \pcc$. From Fig.\ 2 of
\citet{Vazquez07}, it can be seen that the (hydrostatic) cold-phase
density corresponding to this pressure in our simulation is $\sim 150
\pcc$. Since the inflow speed of the colliding streams is 1.22 times the
sound speed in the WNM, the ram pressure is $\sim 1.22^2/\sqrt{5/3} \approx
1$ times the 
thermal pressure,\footnote{Recall the Mach number we use is with respect
to the isothermal sound speed, but the calculation needs to be performed
with the adiabatic one \citep[see also][]{Vazquez06}.} for a total pressure of $P_{\rm tot} = P_{\rm th} +
P_{\rm ram} \approx 10^4~\K \pcc$. The pressure-equilibrium density of the
cold gas at this pressure is seen to be, from that figure, $n \sim 300
\pcc$. 

It is then noteworthy that, although the clump is clearly separated from
the WNM by a sharp boundary (dark green in the top left panels of Figs.\
\ref{fig:clump_structure_center} and \ref{fig:clump_structure}), its
central part (Fig.~\ref{fig:clump_structure_center}) contains densities
ranging from $n \sim 20 \pcc$ to $5000 \pcc$, and is seen to be strongly
turbulent. Specifically, the few-tens-of-Kelvins gas (dark green), which
is mostly associated with the clumps' thin boundaries, is seen to often
extend over much wider regions, penetrating deep into the clump
structure. Presumably, this is gas in transit towards cold-phase
conditions, whose transition has been delayed by the turbulence
\citep{Vazquez03b}. Finally, note that the clump contains a few warmer,
lower-density and lower-pressure ``holes'', which are closer to having
WNM conditions. In summary, the turbulence within the clumps, perhaps
aided by self-gravity { (see below)}, causes strong fluctuations of
about one order of magnitude above and below the canonical steady-state
value of the CNM density. Within this clump, the temperature is $T \sim
20 - 50 \, \K$, whereas the surrounding { warm} gas is still at $T
\sim 5000 \, \K$. The small variability of the temperature within the
clump is consistent with the nearly-isothermal behavior of the gas at
those densities. Indeed, from Fig.\ 2 of \citet{Vazquez07}, it is seen
that the slope $\gamef$ of the $P$ vs. $\rho$ curve for the dense gas is
$\sim 0.8$, while an isothermal behavior would correspond to $\gamef
=1$.  }

The thermal pressure field is { similarly}
seen to have large fluctuations within the clump, in fact
exceeding those seen in the surrounding WNM. The largest values of the
thermal pressure inside the clump are probably caused by the beginning
of the gravitationally-contracting phase of these regions. 
Such large
thermal pressure fluctuations are a reflection of the turbulent 
character of the ram pressure in the clump's environment.



The velocity dispersion within this clump is $\sim 0.7
\kms$. Estimating a sound speed of $\sim 0.4 \kms$ for the gas in the
clump, the implied Mach number is $\sim 1.75$. 
{ Given the slightly softer-than-isothermal
equation of state implies a slightly larger-than-isothermal density jump
in this gas, so that, under the effect of turbulence alone, one should
expect a density contrast $\sim 5$.  The additional density enhancement
{ may be} attributed to incipient gravitational contraction. {Indeed,
the mass of this clump is $\sim 280 M_\odot$, while the Jeans mass for
density $n = 200 \pcc$ is $150 M_\odot$,
indicating that the clump is undergoing gravitational contraction, also
indicated by the fact that it has already formed sink particles.  This is
further illustrated in Fig.\ \ref{fig:MJ_vs_n}, which shows the mass $M$
and the number of Jeans masses $N_{\rm J}$ of the gas above a given
density in the region shown in the rightmost panel of Fig.\
\ref{fig:clump_evol} and in Fig.\ \ref{fig:clump_structure_center}. It
is clearly seen that the clump as a whole is gravitationally unstable,
containing a couple of Jeans masses. The highest density regions do not
appear Jeans unstable, but this may be a consequence of the fact that
this clump has already formed sinks, so the sink mass has already been
removed from the gas phase.} On the
other hand, the lowest-density gas within the clump ($n \sim 20$--$30
\pcc$) is probably due to interference between the condensation process
and the turbulence, causing some accreting gas from the WNM to not be
able to immediately undergo the phase transition to the cold phase.}

\bef
\showone{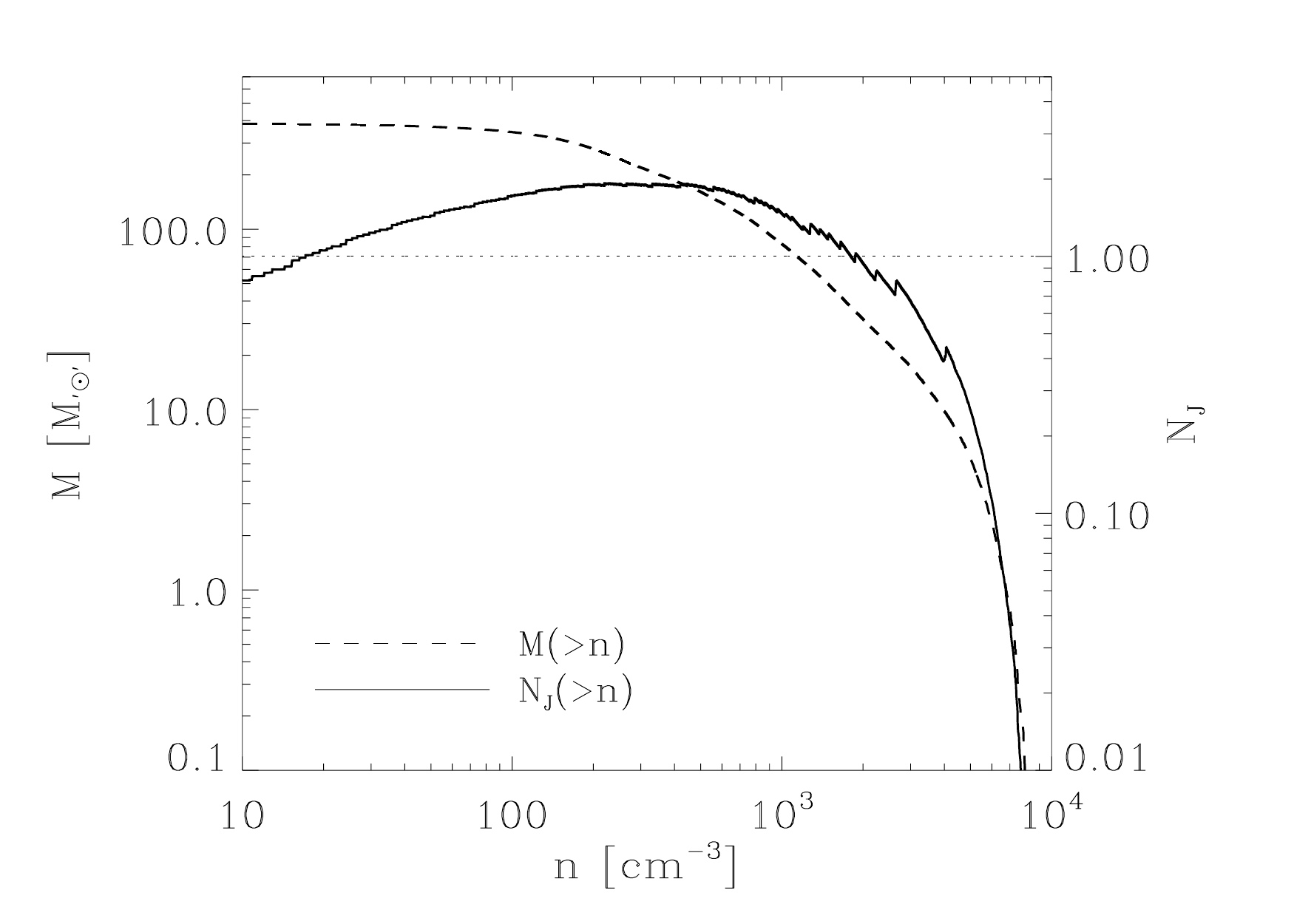}
\caption{Mass $M$ and number of Jeans masses $N_{\rm J}$ of the gas
  above a given density in the region shown in the rightmost panel of
  Fig.\ \ref{fig:clump_evol} and in Fig.\
  \ref{fig:clump_structure_center} ($t = 22.5\, \Myr$). The clump is
  seen to be globally gravitationally unstable (i.e., from its largest
  scales, which correspond to the lowest mean densities.}
\label{fig:MJ_vs_n}
\eef



\bef
\showone{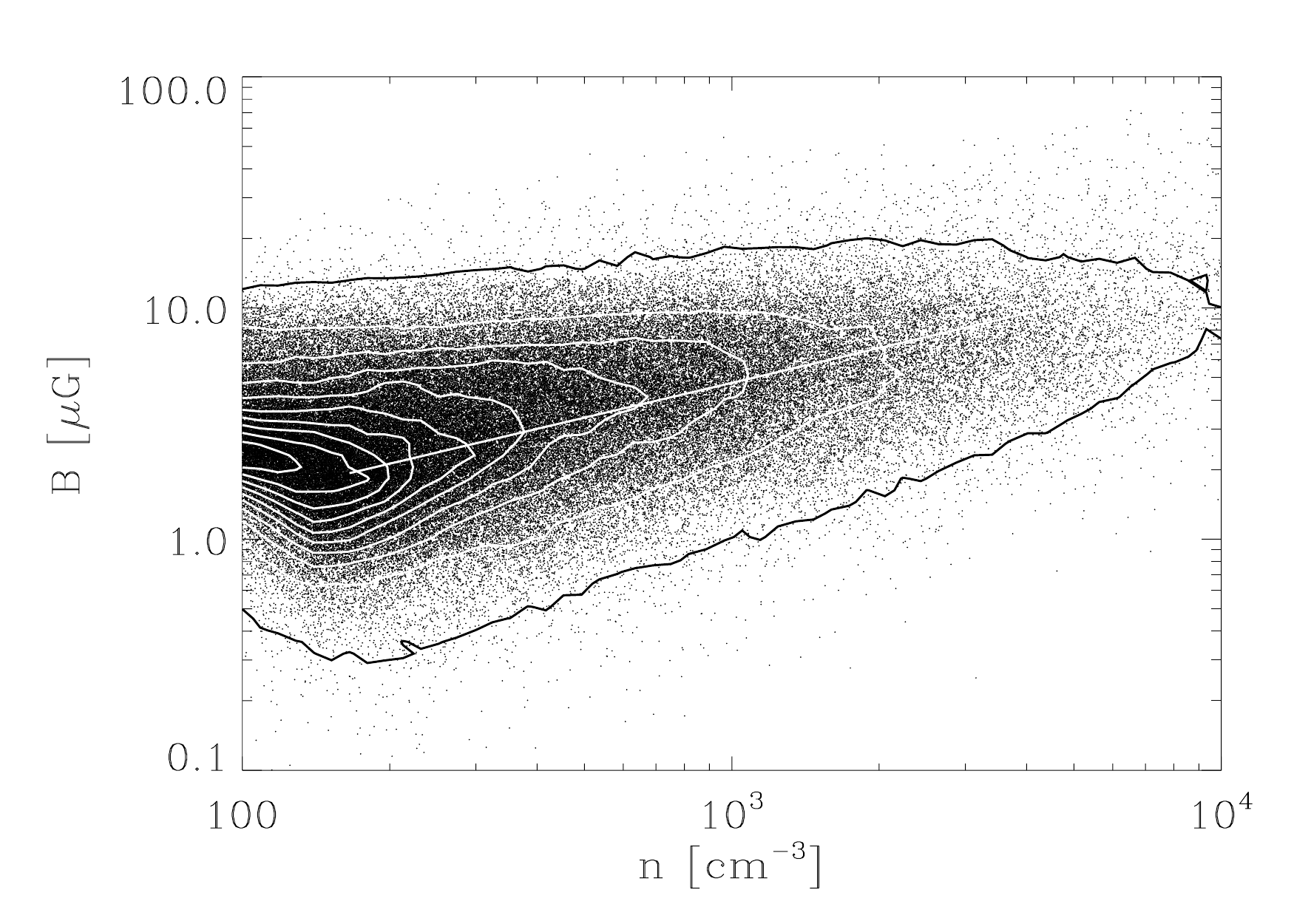}
\caption{{ Scatter plot and two-dimensional histogram showing the
    magnetic field strength and density of all grid points in the
    simulation at $t = 22.5\, \Myr$. The most probable value of the magnetic field at a
    given density (indicated by the locus of the rightmost apexes of
    the contours) is seen to scale with density roughly as $n^{1/2}$
    (straight line), although a large scatter of almost two orders of
    magnitude is seen around this mean trend at each density.} On the
  other hand, the maximum magnetic field strength scales only weakly
  with the gas density (i.e., $B_{\sm{max}} \propto n^{0.15}$ for the
  uppermost contour line).}
\label{fig:b-n-scatter}
\eef

\bef
\showone{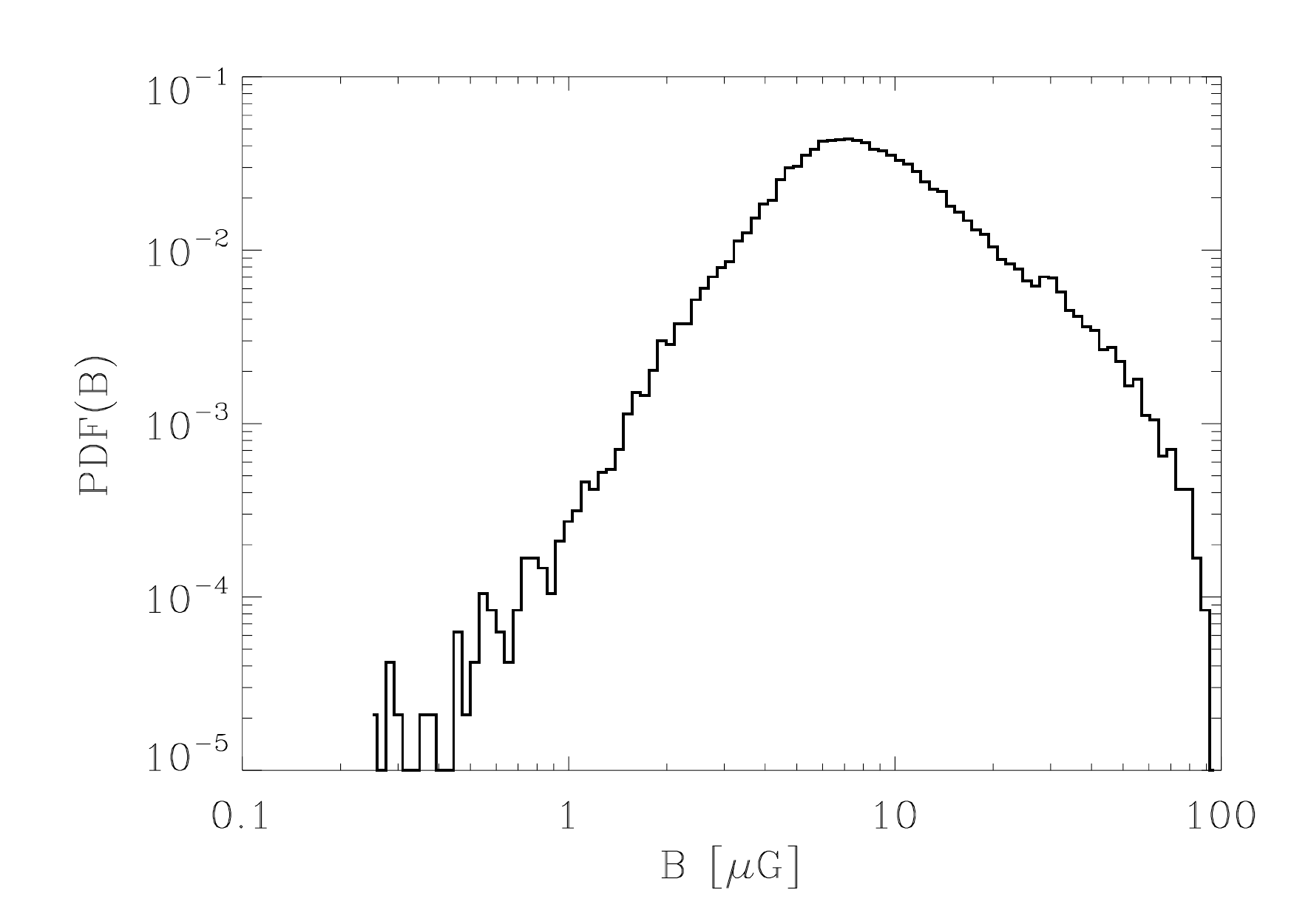}
\caption{Probability density function of the magnetic field
  strength in the high density regions ($n \ge 5\times 10^3 \,
  \pcc$), showing a variability of one order of magnitude of the
field strength above and below a most probable value of $\langle B \rangle
\sim 6~\mu$G at $t = 22.5\, \Myr$. 
}
\label{fig:bmag_pdf}
\eef

Concerning the magnetic field, from Figs.\
\ref{fig:clump_structure_center} and \ref{fig:clump_structure}, we see that it tends to be aligned with the
velocity field in the dense regions, although it is also highly
distorted there (recall that the initial configuration had the magnetic field
parallel to the $x$ axis), a phenomenon already observed in the 2D
simulations of \citet{Passot95}. { The visual impression of alignment
is confirmed by the histogram of $\vv \cdot \vB/|\vv| |\vB|$, which is 
shown in Fig.\ \ref{fig:v_dot_B_hist} for all of the dense ($n > 5
\times 10^3 \pcc$) gas in the simulation. The histogram clearly exhibits
peaks at 1 and $-1$, indicating alignment.}  The tangling { in the
clumps} indicates that the field has been strongly distorted by the
turbulent motions in the compressed regions. In this case, the alignment
with the velocity field is probably a consequence of the fact that
motions non-parallel to the field tend to stretch and align it along
with them \citep{Hennebelle00}. { A similar alignment is observed in
runs with a more realistic mean field strength of $3 \mu$G, and
including an initial fluctuating component of the field, suggesting that
this result may actually be expected to apply in actual interstellar
clouds. }

\bef
\showone{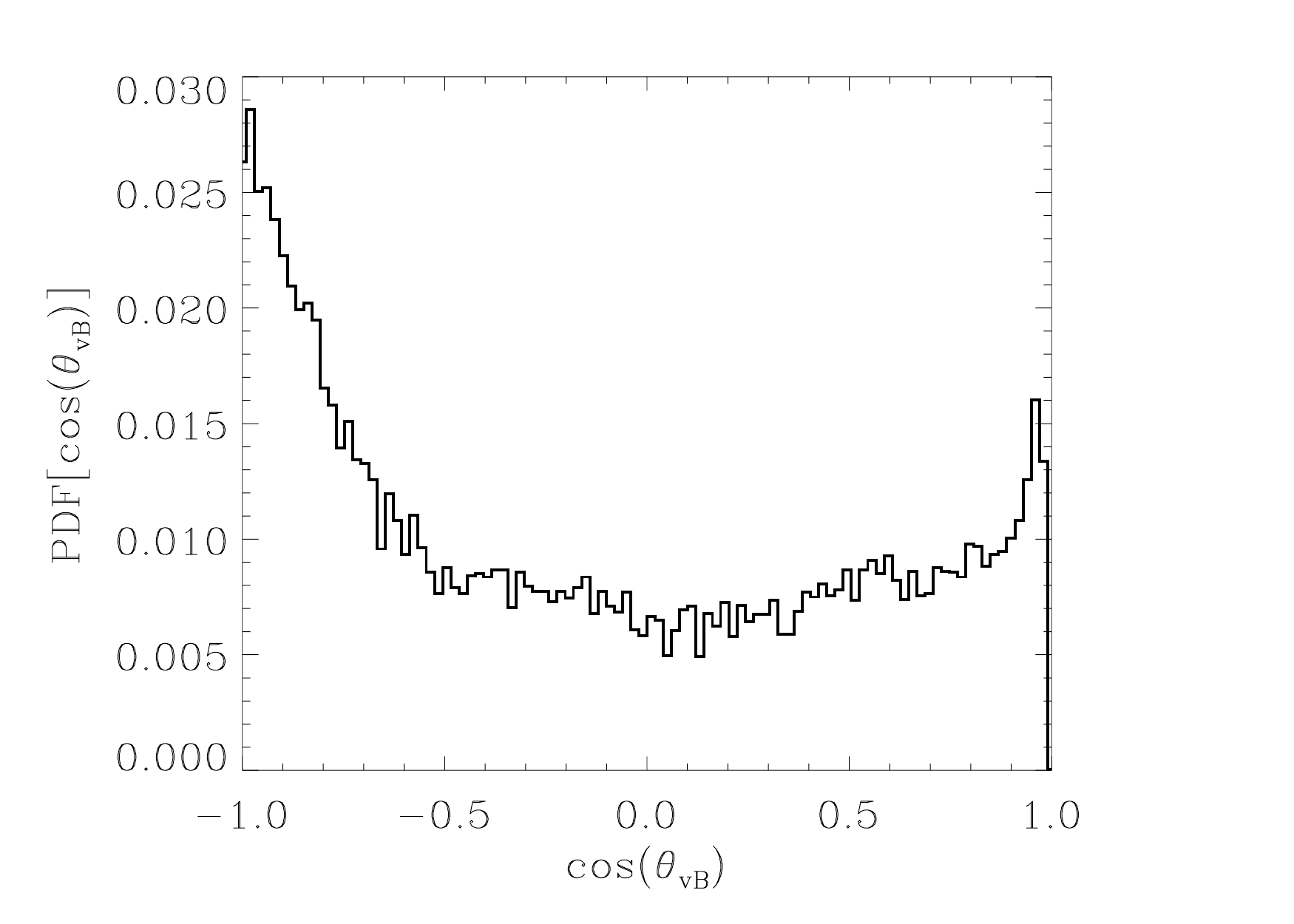}
\caption{Histogram of $\cos\theta_{vB} = \vv \cdot \vB/vB$ computed
over all of the high-density ($n > 5 \times 10^3 \pcc$) gas in the
simulation, where $v$ and $B$ are respectively the magnitudes of the
vectors $\vv$ and $\vB$ at $t = 22.5\, \Myr$. The velocity and magnetic
field vectors are clearly seen to show a strong tendency to be either
parallel or antiparallel.}
\label{fig:v_dot_B_hist}
\eef

 Fig.~\ref{fig:b-n-scatter} shows a scatter plot ({\it dots}) and
  a two-dimensional histogram ({\it contours}) of the magnetic field
  strength versus the density for each pixel in the simulation. One
  can see that, from $n \sim 100 \pcc$ to $10^4 \pcc$, the most
  probable value of the magnetic field, indicated by the locus of the
  rightmost apexes of the contours, appears to scale roughly as
  $n^{1/2}$, although with great scatter around this value.


It is worth noting that at densities $n \sim 10^4 \pcc$, the
mean field strength is $\langle B \rangle \sim 10
\mu$G, although field strengths ranging from $\sim 1$ to $\sim 100
\mu$G are observed. { This is illustrated in Fig.~\ref{fig:bmag_pdf}, which
shows the field strength distribution in the high
  density gas only ($n \ge 5\times 10^3 \pcc$).} This suggests that
strongly as well as weakly magnetized cores should exist
within the clumps, implying that some of the non-detections of the field
through, for example, Zeeman measurements \citep{Crutcher99, Crutcher04},
should actually correspond to very low field strengths, rather than to
alignment effects that mask the field.  




\bef
\showonep{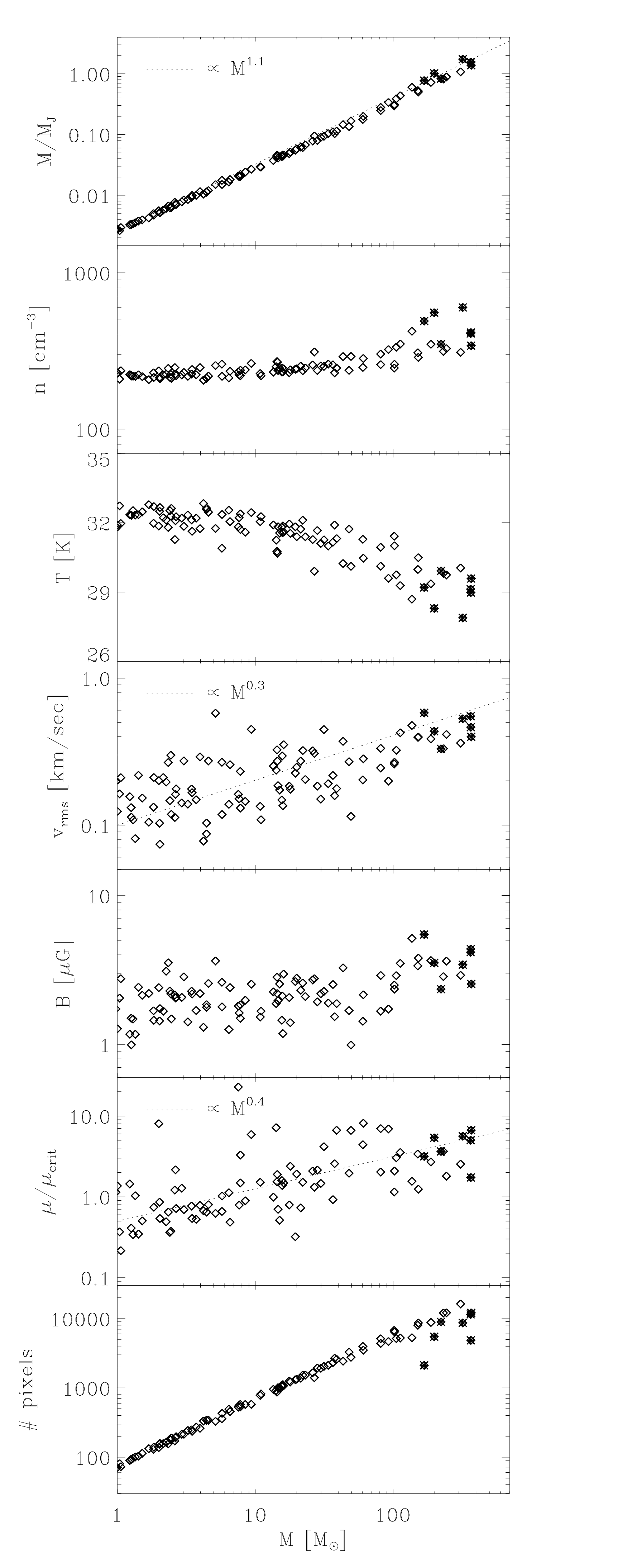}
        \caption{Statistical properties of the molecular clumps
          identified in the cloud at $t = 22.5 \, \Myr$. Here, we
          { define} clumps as connected regions with { densities
          above} $200\, \cm^{-3}$. These quantities are
          plotted as functions of the total mass in the clumps
          (i.e. $M = M_{\sm{gas}} + M_{\sm{sinks}}$). The clumps
          indicated by { asterisks} (*) have already formed sink
          particles. $M_J$ is the thermal Jeans mass and $\mu$ is the
          mass-to-flux ratio, where $\mu_{\sm{crit}}$ is the critical
          value. We also show the best fit power laws as indicated in
          the panels.}
\label{fig:clump_statistics}
\eef

\subsubsection{Statistics of mean clump properties} \label{sec:collective}

In order to study the collective properties of all clumps in the
molecular cloud, we apply a simple algorithm to identify them in our
simulation data.  We define clumps as connected regions with density
above a certain threshold. For comparison, we use two threshold
values: $n_{\sm{thres}} = 200 \, \cm^{-3}$ and $n_{\sm{thres}} = 500
\, \cm^{-3}$. { We choose these values because they bracket the
steady-state value of the density of the CNM (cf.\ sec.\
\ref{sec:int_struc}). We have found that the
clump statistics are fairly insensitive to the different density
thresholds, suggesting that our results are not significantly biased by
our choice of threshold.}

In Fig.~\ref{fig:clump_statistics} we show some of the averaged internal
properties of the clumps found { in the whole cloud} at $t = 22.5 \,
\Myr$. The masses of these clumps span the range $2 - 400 \, \Msol$ at
this time. Clumps more massive than $\sim 200 \, \Msol$ are close to
being Jeans unstable (see the top panel of this figure, where we show
the ratio $M/M_J$ as a function of the clump mass $M$). Some regions
within these massive clumps are collapsing and will form stars. It is
interesting, however, that the {self-gravitating} clumps as a whole
tend to have values of $M/M_{\rm J}$ {not much larger than
unity. At first sight, this might appear contradictory with the notion
that molecular clouds contain many Jeans masses. However, this apparent
contradiction is resolved by noting that our entire cloud, formed of
many clumps, does contain a large number of Jeans masses, as illustrated
in Fig.\ \ref{fig:M_MJ_whole_cloud}, which shows the total mass of the
cloud (defined as gas with density $n > 100\, \pcc$) and the ratio of this
mass to its Jeans mass, clearly indicating 
that, by the end of the simulation, the entire cloud contains roughly 100
Jeans masses. This supports the notion that giant molecular clouds may
consist of molecular clumps immersed in an atomic substrate
\citep[e.g.,][]{Goldsmith05, Hennebelle06}.}

\bef
\showone{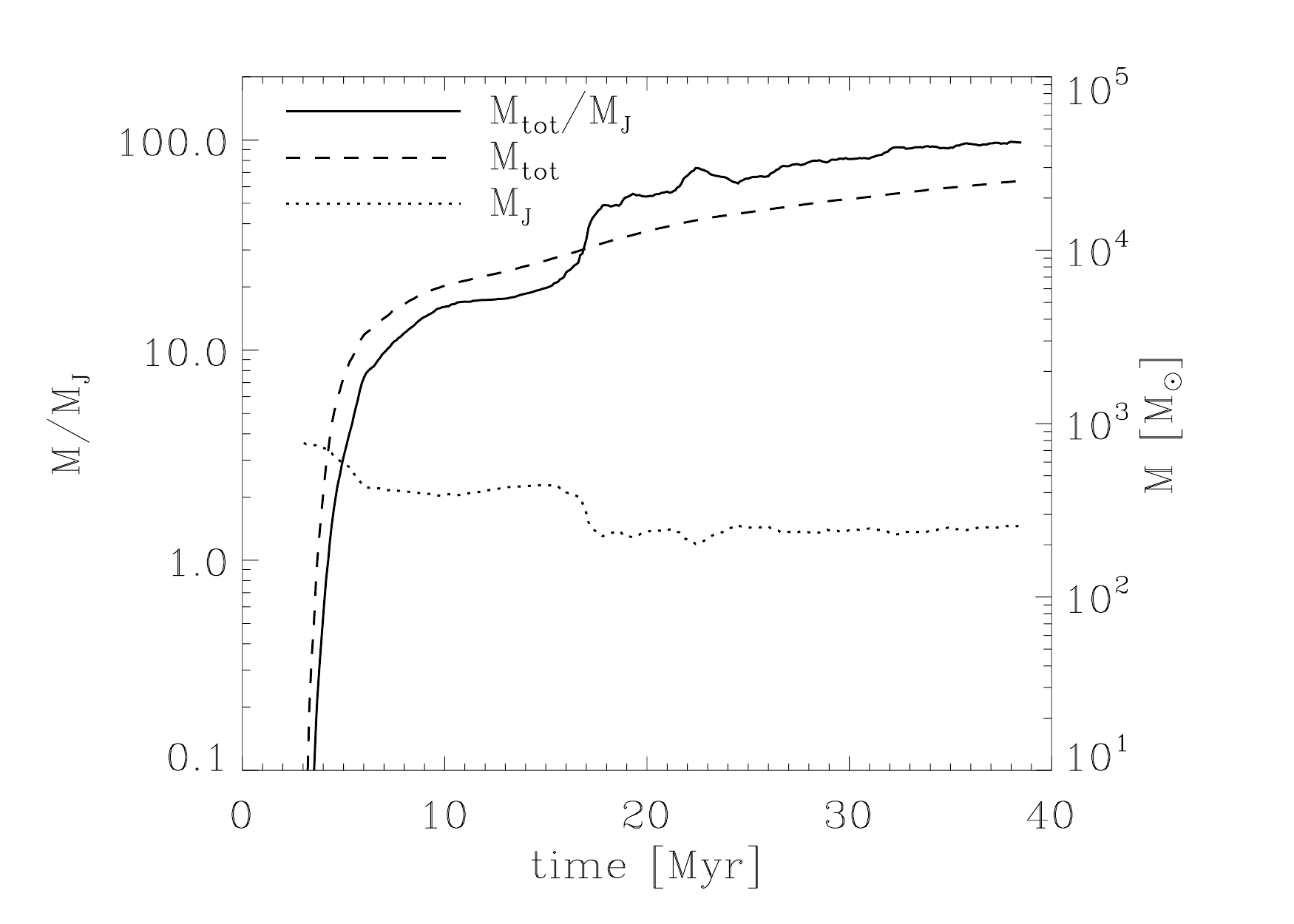}
\caption{Evolution of the total cloud mass (defined as gas with density
$n > 100\, \pcc$) and ratio of this mass to the cloud's Jeans mass,
showing that the entire cloud rapidly grows to contain a large number of
Jeans masses.}
\label{fig:M_MJ_whole_cloud}
\eef

The average density and temperature depend only weakly on the clump
masses. This reflects the fact that { the filling factor of the
higher-density gas within the clumps is very low (i.e., most of the mass
is at the lower densities), and thus the mean density of the clumps is
always very close to the threshold density for defining them, with the
most massive clumps having slightly larger mean densities
\citep{Vazquez97}.} 
The near constancy of the clumps'
densities and temperatures is also indicated by the almost linear
relation between the clumps' masses and their ratio of mass to Jeans
mass, as shown in Fig.~\ref{fig:clump_statistics}.

The velocity dispersion $\sigma$ in the clumps increases with clump
mass, which, together with the fact that the clumps have roughly the
same density, implies that $\sigma$ increases with size. However, note
that the dynamical range of the clump sizes is { less} than one
decade and the scatter in the velocity data is quite large. This
prevents us from fitting a proper $\sigma$-$L$ relation {
  here}. Estimating a { typical sound speed} of $\sim 0.4 \, \km\,
\sec^{-1}$ we find that most of the clumps are sub- or transonic. Only
the most massive clumps, which are already in the state of collapse,
develop larger supersonic velocities, { suggesting that such
  velocities are the result of the gravitational collapse of the
  clumps, or regions within them (the cores)}.

The fact that the turbulence in the clumps shortly after their 
condensation is transonic, just as is the turbulence in the
surrounding WNM, is remarkable. It appears to be consequence that the
ram pressure fluctuations in the cold gas are excited by the ram
pressure of the warm gas, i.e., $\rho_{\sm{w}} v_{\sm{w}}^2 \sim
\rho_{\sm{c}} v_{\sm{c}}^2$.  { In addition,} the cold, dense
clumps evolve quickly into thermal-{ pressure} equilibrium with their
warm surroundings (due to the thermal bistability), so that $\rho_{\sm{w}}
c_{\sm{w}}^2 \sim
\rho_{\sm{c}} c_{\sm{c}}^2$. The two conditions combined imply that
the thermal Mach numbers in both media are comparable (i.e., $M_{\sm{w}}
\sim M_{\sm{c}}$).

The typical mean field strength in the clumps { exhibits an
interesting dichotomy: clumps with $M \lesssim 100 \Msol$ have a mean
field strength $B \sim 2~\mu$G, independent of the clump's mass, while
clumps with $M \gtrsim 100 \Msol$, of which more than half have already
formed sinks, have systematically larger field strengths of nearly twice
that amount. This result is in agreement with observational results that
the field strength is essentially independent of gas density for the WNM
and CNM, and only begins to increase with density at higher densities,
where presumably gravitational contraction is at work
\citep[e.g.][]{Crutcher03, Heiles05}. It is nevertheless interesting
that the typical field strength for the low-mass clumps is roughly twice
the mean value for the whole simulation, indicating that some amount of
mean field amplification exists in the clumps with respect to the WNM, even
if by only a factor of $\sim 2$.}


Finally, we note that almost all clumps with mass above $10\, \Msol$
show a critical or 
super-critical mass-to-flux ratio, $\mu$, with a mass dependence of $\mu
\propto M^{0.25 - 0.4}$,\footnote{We use the projected area of the clump,
$A_{\sm{yz}}$, and the averaged normal field component, $\langle
B_x\rangle = V^{-1}\,\int\,\di V\,B_x$, where $V$ is the volume of the
clump, to calculate the mass-to-flux ratio, i.e. $\mu =
M_{\sm{clump}}/A_{\sm{yz}}\langle B_x\rangle$.} although this
result may 
{ be an artifact of the low degree of magnetization of our
simulations} (recall the mean field of our magnetized
simulation is $1 \mu$G).

\bef
\showone{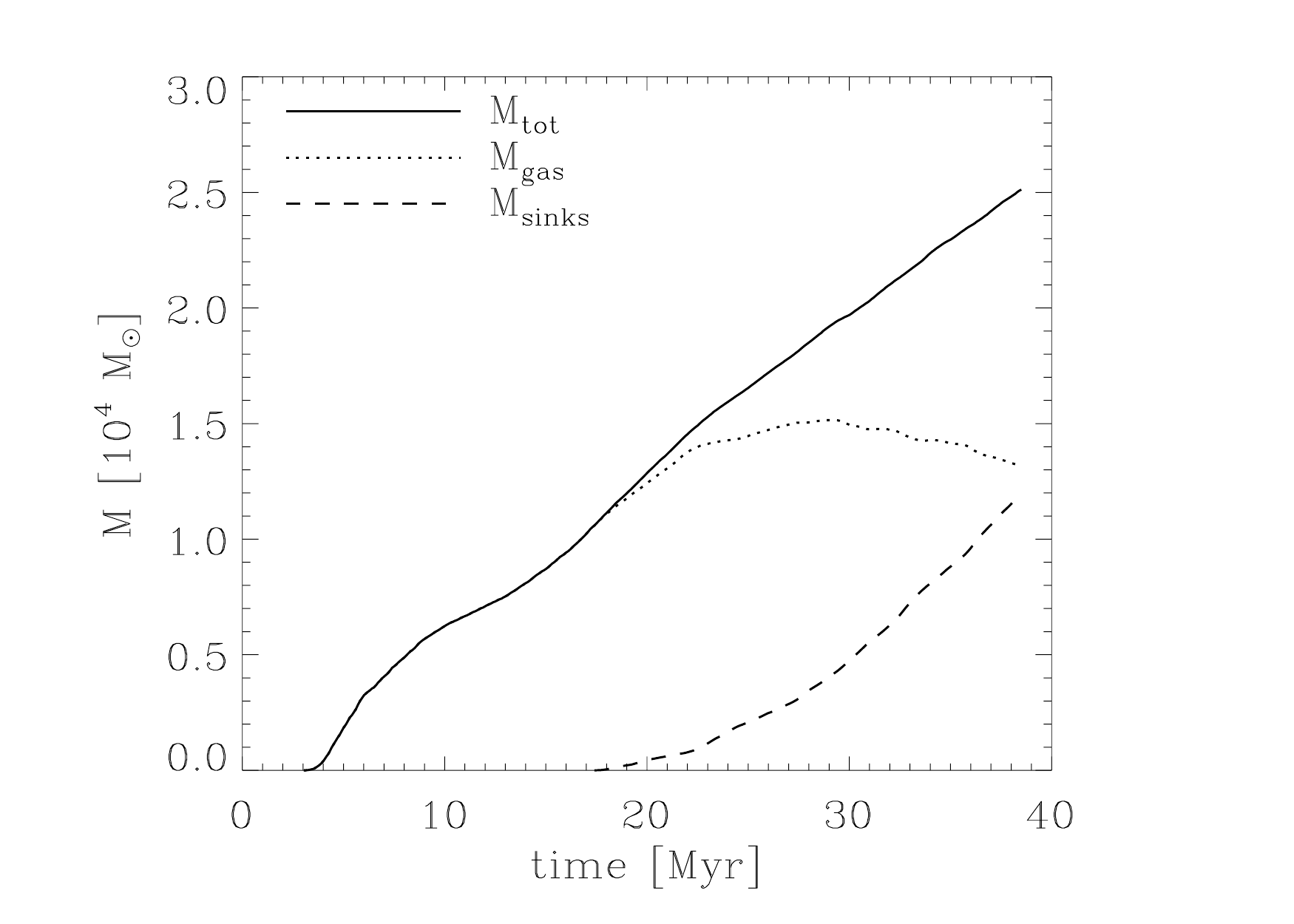}
\caption{Mass evolution of the dense gas (i.e. $n > 100 \, \cm^{-3}$,
  M$_{\sm{gas}}$), sink particles (M$_{\sm{sinks}}$,) and their
  sum (M$_{\sm{tot}}$). The mass accretion rate of the
  collapsing regions increases from $10^{-4}$ to $10^{-3}\, \Msol\,
  \yr^{-1}$ during the cloud evolution. Note that our model does not include feedback effects which should ultimately limit the star formation efficiency~\citep[e.g. see][for an estimate of the time at which the clould could be disrupted by the formed OB stars.]{Vazquez07}
}
\label{fig:mass_evol}
\eef

\subsubsection{Evolution of the cloud} \label{sec:cloud_evol}

While the individual clumps inside the molecular cloud grow and merge,
the cloud continues to accrete mass from the WNM (see
Fig.~\ref{fig:mass_evol} for the cloud mass 
evolution). Eventually, this leads to the global contraction of the
entire cloud. In our simulation this happens at $t \sim 20\, \Myr$
which is about $15\, \Myr$ after the first dense clumps have
formed. The increased gravitational potential in the centre of the
cloud further compresses the gas, therefore converting an increasing
amount of diffuse gas into the dense phase.
This relieves concerns that the accumulation length required to attain
molecular cloud-like column densities from one-dimensional
accumulation of WNM gas is exceedingly long \citep[$\sim 1$ kpc;
][sec. 2.3]{McKee07}, since much of the column density increase is
provided by the three-dimensional gravitational compression of the
gas. The importance of lateral gravitational contraction was already
pointed out in~\citet{Hartmann01} and was recently also confirmed in
three dimensional simulations by~\citet{Heitsch08c} \citep[see
also][]{Dobbs08}. The one-dimensional compression only provides the
cooling and compression necessary for self-gravity to become
important, which then provides the remaining necessary compression
\citep[see also][]{Elmegreen07}.

The late global contraction finally leads to enhanced gas densities
and large ($\sim 10\, \pc$), dense, coherent, and collapsing regions of mainly
``molecular'' gas, in which local collapse events occur before the
global collapse is completed \citep[e.g.][]{Klessen01, MacLow04}. Eventually, a high-density and high-velocity dispersion region forms in the overall minimum of gravitational potential, when the global collapse finally reaches center. We expect that this stage may result in the conditions where massive stars can form~\citep[e.g.,][]{ZinneckerYorke07, Vazquez08, Vazquez09}.


The global evolution of our fiducial magnetic simulation, with a mean
field strength of $1 \mu$G, is in general similar to the evolution of
the non-magnetic fiducial simulation in \citet[][run L256$\Delta
v0.17$ there]{Vazquez07}, in the sense that the collision of the WNM
flows produces a CNM cloud that, due to the combined action of
compression and cooling, becomes gravitationally unstable and begins
to contract and undergo collapse at localized sites. Also similar are
the evolutions of the dense-gas and stellar masses \citep[compare
Fig.\ \ref{fig:mass_evol} to fig. 5 of][]{Vazquez07}, and the
formation of a ring at the periphery of the cloud~\citep{Burkert04}. Subtle differences,
however, do exist beteen the two simulations, due mainly to the
presence of the magnetic field, albeit weak, and to the slightly
weaker inflow speed of our simulation compared to that of run
L256$\Delta v0.17$. In particular, in the present simulation, the
global collapse of the ring occurs at a significantly later time than
in run L256$\Delta v0.17$ ($t \sim 40$ Myr vs. $t \sim 23$ Myr,
respectively). This is possibly a consequence of the lower inflow
speed, which does not cause as strong an ejection of material in the
radial direction, as well as of the presence of the magnetic field,
which is perpendicular to this ejection direction, both allowing a
greater concentration of material in the central parts of the cloud,
and a lower mass of the ring. As a consequence, in our simulation the
first local collapse events occur in the central parts of the cloud,
while in run L256$\Delta v0.17$ they occurred in the peripheral ring.

\section{Summary and Conclusions}
\label{sec:conclusions}

In this paper we have reported results on the physical properties of
the dense gas (which we refer to as ``molecular'') structures formed by
transonic compressions in the diffuse atomic medium, using 3D MHD
simulations including self-gravity, and  radiative heating and cooling
laws leading to thermal bistability  of the gas. We have defined the
clumps as connected regions with densities $n \geq 200$ cm$^{-3}$, which
selects the clumps formed by a phase transition from the diffuse to the
dense phases of thermal instability (TI). We do not consider in our
statistics the substructures within these clumps, which would correspond
to dense molecular cores.

The ram pressure from the accretion of WNM gas into the clumps
contributes a net ram pressure, in addition to the thermal pressure of
the WNM, causing the clumps' densities to overshoot past the typical
conditions of the CNM ($n \sim 50 \pcc$), well into the realm of physical
conditions typical of { large} molecular clouds ($n \gtrsim 200
\pcc$). Moreover, since the ram pressure 
from the diffuse medium is turbulent and fluctuating, it induces transonic
turbulence within the clumps which, as a consequence of the joint
conditions of ram and thermal pressure balance, must have an rms Mach
number comparable to that in the diffuse gas.
The transonic turbulence in the clumps induces significant density
fluctuations, which then provide the seeds for
subsequent local gravitational collapse as the clumps approach
their Jeans mass. 

We found that the transition between such clumps and the diffuse medium
is generally sharp, with both media being at roughly the same thermal
pressure, similarly to the situation in the classical two-phase medium
of \citet{Field69}. However, the clumps contain large density
fluctuations within them, of up to one order of magnitude above and
below the nominal pressure-equilibrium density value, caused by the
presence of thermally unstable gas still in transit towards the cold
phase on the one hand, and to local gravitational contraction on the
other. Thus, the boundaries of the clumps, which generally consist of
thin layers of thermally unstable gas, often become extended and
penetrate deep into the clumps. The clumps are nearly isothermal
inside, with temperatures in the range $\sim 20$--50 K, as
consequence of the density fluctuations within the clumps and the nearly
isothermal equation of state governing the high-density gas.

Another key difference between the classical model and the results of
our simulations is that the clumps are formed {\it dynamically} {by
the compressions in the surrounding WNM}, implying that they are subject
to continuous accretion from the WNM {driven by its ram
pressure \citep{Ballesteros99a, Ballesteros06}. This in turn causes}
the clumps' mass and
size to grow in time. Thus, the clumps'
boundaries are {ram-pressure-driven} {\it phase-transition fronts} 
and clump growth occurs mainly by accretion through
their boundaries, rather than by coagulation, as was the case in earlier
models of the ISM \citep[e.g., ][]{Kwan83}. In turn, this mass flux
drives the clumps to 
eventually become gravitationally unstable and collapse \citep[see
also][for an analogous situation in isothermal flows, with clumps being
bounded by accretion shocks]{Gomez07}. 

The magnetic field shows a significant level of alignment with the
velocity field, but also large fluctuations in magnitude and direction
{ inside the clumps}, suggesting that it has been significantly
distorted by the turbulent motions in the dense gas. 
We also find very similar distortions of the magnetic field structure
by turbulent motions in the case with larger a initial field strengths
of $3\, \mu\G$. This suggests that gas streams and field lines are
likely to be aligned in the cases of { either} a weak { or} a
strong magnetic field.

The molecular clumps and the cloud as a whole are dynamical and evolve
with time, with important consequences for their ability to form stars
\citep[e.g.,][]{MacLow04, Ballesteros07}. After some 20 Myr of
evolution, some regions have already undergone local collapse and
started to form stars, while other clumps do not yet show signs of
star formation, similarly to the suggestion by \citet{Elmegreen07} for
clouds behind the spiral arms of the Galaxy. From Fig. 14, we see that
by $t \sim 28$ Myr, roughly 15-20\% of the total mass in the cloud
(dense gas + sinks) has been converted to sinks. By this time,
according to the estimates of \citet{Vazquez07}, based on the
prescription by \citet{Franco94}, the cloud could be destroyed by the
newly formed massive stars. Since SF began in the simulation at $t
\sim 17$ Myr, this implies that the stellar age spread in the {\it
  entire cloud} should be $\sim 10$ Myr. Note, however, that our
entire cloud, with a physical size of $\sim 80$ pc is analogous to a
giant cloud complex, rather than to an isolated cloud. Local, isolated
SF sites of sizes $\sim 10$ pc, have smaller age spreads.

During the evolution of the cloud, global gravitational focusing
enlarges connected molecular regions in the centre of the
cloud~\citep[see also][]{Burkert04, Hartmann07}. At the stage when the
global contraction reached the center of the cloud, we expect that the
conditions should be reached where massive stars could form\citep[see,
e.g., ][]{ZinneckerYorke07, Vazquez08, Vazquez09}.

We conclude that the formation and evolution of clumps in a thermally
bistable medium is a highly complex process that retains some of the
features from the classical two-phase model, such as the frequent
presence of sharp density discontinuities, which separate the cold
clumps from the warm diffuse medium, while at the same time exhibiting
a much more complex structure, consisting of an intrincate filament
network connecting the clumps, and made up of mainly thermally
unstable gas. Furthermore, the clumps are internally turbulent, and
thus have density fluctuations of up to one order of magnitude even
before they become gravitationally unstable. The role of the magnetic
field appears negligible at the relatively low magnetization levels we
have considered in this paper. In a subsequent paper, we will consider
more strongly magnetized regimes, including ambipolar diffusion.

\section*{Acknowledgements}

We thank the anonymous referee for useful comments and suggestions on
our work which helped to improve this paper. The FLASH code was
developed in part by the DOE-supported Alliances Center for
Astrophysical Thermonuclear Flashes (ASC) at the University of
Chicago. Our simulations were carried out on the Cluster Platform 4000
(KanBalam) at DGSCA-UNAM and at HLRB II of the Leibniz Rechenzentrum,
Garching.  RB is funded by the DFG under the grant BA
3607/1-1. E.V.-S. acknowledges financial support from CONACYT grant
U47366-F.  R.S.K. thanks for subsidies from the German Science
Foundation (DFG) under Emmy Noether grant KL 1358/1 and grants KL
1358/4 and KL 1358/5. This work was supported in part by a FRONTIER
grant of Heidelberg University sponsored by the German Excellence
Initiative, as well as by the Federal Ministry of Education and
Research via grant 17EcCZXd.

\input{HMP_astroph_v2.bbl}

\end{document}

%% file: journals.tex
\def\jnl@style#1{{\rmfamily#1}}%
\def\jref@jnl#1{{\jnl@style#1}}%

\newcommand\aj{\jref@jnl{AJ}}%
\newcommand\araa{\jref@jnl{ARA\&A}}%
\newcommand\apj{\jref@jnl{ApJ}}%
\newcommand\apjl{\jref@jnl{ApJ}}%
\newcommand\apjs{\jref@jnl{ApJS}}%
\newcommand\ao{\jref@jnl{Appl.~Opt.}}%
\newcommand\apss{\jref@jnl{Ap\&SS}}%
\newcommand\aap{\jref@jnl{A\&A}}%
\newcommand\aapr{\jref@jnl{A\&A~Rev.}}%
\newcommand\aaps{\jref@jnl{A\&AS}}%
\newcommand\azh{\jref@jnl{AZh}}%
\newcommand\baas{\jref@jnl{BAAS}}%
\newcommand\jrasc{\jref@jnl{JRASC}}%
\newcommand\memras{\jref@jnl{MmRAS}}%
\newcommand\mnras{\jref@jnl{MNRAS}}%
\newcommand\pra{\jref@jnl{Phys.~Rev.~A}}%
\newcommand\prb{\jref@jnl{Phys.~Rev.~B}}%
\newcommand\prc{\jref@jnl{Phys.~Rev.~C}}%
\newcommand\prd{\jref@jnl{Phys.~Rev.~D}}%
\newcommand\pre{\jref@jnl{Phys.~Rev.~E}}%
\newcommand\prl{\jref@jnl{Phys.~Rev.~Lett.}}%
\newcommand\pasp{\jref@jnl{PASP}}%
\newcommand\pasj{\jref@jnl{PASJ}}%
\newcommand\qjras{\jref@jnl{QJRAS}}%
\newcommand\skytel{\jref@jnl{S\&T}}%
\newcommand\solphys{\jref@jnl{Sol.~Phys.}}%
\newcommand\sovast{\jref@jnl{Soviet~Ast.}}%
\newcommand\ssr{\jref@jnl{Space~Sci.~Rev.}}%
\newcommand\zap{\jref@jnl{ZAp}}%
\newcommand\nat{\jref@jnl{Nature}}%
\newcommand\iaucirc{\jref@jnl{IAU~Circ.}}%
\newcommand\aplett{\jref@jnl{Astrophys.~Lett.}}%
\newcommand\apspr{\jref@jnl{Astrophys.~Space~Phys.~Res.}}%
\newcommand\bain{\jref@jnl{Bull.~Astron.~Inst.~Netherlands}}%
\newcommand\fcp{\jref@jnl{Fund.~Cosmic~Phys.}}%
\newcommand\gca{\jref@jnl{Geochim.~Cosmochim.~Acta}}%
\newcommand\grl{\jref@jnl{Geophys.~Res.~Lett.}}%
\newcommand\jcp{\jref@jnl{J.~Chem.~Phys.}}%
\newcommand\jgr{\jref@jnl{J.~Geophys.~Res.}}%
\newcommand\jqsrt{\jref@jnl{J.~Quant.~Spec.~Radiat.~Transf.}}%
\newcommand\memsai{\jref@jnl{Mem.~Soc.~Astron.~Italiana}}%
\newcommand\nphysa{\jref@jnl{Nucl.~Phys.~A}}%
\newcommand\physrep{\jref@jnl{Phys.~Rep.}}%
\newcommand\physscr{\jref@jnl{Phys.~Scr}}%
\newcommand\planss{\jref@jnl{Planet.~Space~Sci.}}%
\newcommand\procspie{\jref@jnl{Proc.~SPIE}}%

%% file: HMP_astroph_v2.bbl
\begin{thebibliography}{}

\bibitem[\protect\citeauthoryear{{Audit} \& {Hennebelle}}{{Audit} \&
  {Hennebelle}}{2005}]{Audit05}
{Audit} E.,  {Hennebelle} P.,  2005, \aap, 433, 1

\bibitem[\protect\citeauthoryear{{Ballesteros-Paredes}}{{Ballesteros-Paredes}}%
{2006}]{Ballesteros06}
{Ballesteros-Paredes} J.,  2006, \mnras, 372, 443

\bibitem[\protect\citeauthoryear{{Ballesteros-Paredes}, {Hartmann} \&
  {V{\'a}zquez-Semadeni}}{{Ballesteros-Paredes} et~al.}{1999}]{Ballesteros99b}
{Ballesteros-Paredes} J.,  {Hartmann} L.,    {V{\'a}zquez-Semadeni} E.,  1999,
  \apj, 527, 285

\bibitem[\protect\citeauthoryear{{Ballesteros-Paredes}, {Klessen}, {Mac Low} \&
  {Vazquez-Semadeni}}{{Ballesteros-Paredes} et~al.}{2007}]{Ballesteros07}
{Ballesteros-Paredes} J.,  {Klessen} R.~S.,  {Mac Low} M.-M.,
  {Vazquez-Semadeni} E.,  2007, in {Reipurth} B.,  {Jewitt} D.,   {Keil} K.,
  eds, Protostars and Planets V {Molecular Cloud Turbulence and Star
  Formation}.
pp 63--80

\bibitem[\protect\citeauthoryear{{Ballesteros-Paredes}, {V{\'a}zquez-Semadeni}
  \& {Scalo}}{{Ballesteros-Paredes} et~al.}{1999}]{Ballesteros99a}
{Ballesteros-Paredes} J.,  {V{\'a}zquez-Semadeni} E.,    {Scalo} J.,  1999,
  \apj, 515, 286

\bibitem[\protect\citeauthoryear{{Bate}, {Bonnell} \& {Price}}{{Bate}
  et~al.}{1995}]{Bate95}
{Bate} M.~R.,  {Bonnell} I.~A.,    {Price} N.~M.,  1995, \mnras, 277, 362

\bibitem[\protect\citeauthoryear{{Beck}}{{Beck}}{2001}]{Beck01}
{Beck} R.,  2001, Space Science Reviews, 99, 243

\bibitem[\protect\citeauthoryear{{Burkert} \& {Hartmann}}{{Burkert} \&
  {Hartmann}}{2004}]{Burkert04}
{Burkert} A.,  {Hartmann} L.,  2004, \apj, 616, 288

\bibitem[\protect\citeauthoryear{{Crutcher}, {Heiles} \& {Troland}}{{Crutcher}
  et~al.}{2003}]{Crutcher03}
{Crutcher} R.,  {Heiles} C.,    {Troland} T.,  2003, in {Falgarone} E.,
  {Passot} T.,  eds, Turbulence and Magnetic Fields in Astrophysics Vol.~614 of
  Lecture Notes in Physics, Berlin Springer Verlag, {Observations of
  Interstellar Magnetic Fields}.
pp 155--181

\bibitem[\protect\citeauthoryear{{Crutcher}}{{Crutcher}}{2004}]{Crutcher04}
{Crutcher} R.~M.,  2004, \apss, 292, 225

\bibitem[\protect\citeauthoryear{{Crutcher}, {Troland}, {Lazareff}, {Paubert}
  \& {Kaz{\` e}s}}{{Crutcher} et~al.}{1999}]{Crutcher99}
{Crutcher} R.~M.,  {Troland} T.~H.,  {Lazareff} B.,  {Paubert} G.,    {Kaz{\`
  e}s} I.,  1999, \apjl, 514, L121

\bibitem[\protect\citeauthoryear{{de Avillez} \& {Breitschwerdt}}{{de Avillez}
  \& {Breitschwerdt}}{2004}]{Avillez04}
{de Avillez} M.~A.,  {Breitschwerdt} D.,  2004, \aap, 425, 899

\bibitem[\protect\citeauthoryear{{Dobbs}, {Glover}, {Clark} \&
  {Klessen}}{{Dobbs} et~al.}{2008}]{Dobbs08}
{Dobbs} C.~L.,  {Glover} S.~C.~O.,  {Clark} P.~C.,    {Klessen} R.~S.,  2008,
  \mnras, 389, 1097

\bibitem[\protect\citeauthoryear{{Duffin} \& {Pudritz}}{{Duffin} \&
  {Pudritz}}{2008}]{Duffin08}
{Duffin} D.~F.,  {Pudritz} R.~E.,  2008, \mnras\ in press

\bibitem[\protect\citeauthoryear{{Elmegreen}}{{Elmegreen}}{2007}]{Elmegreen07}
{Elmegreen} B.~G.,  2007, \apj, 668, 1064

\bibitem[\protect\citeauthoryear{{Field}}{{Field}}{1965}]{Field65}
{Field} G.~B.,  1965, \apj, 142, 531

\bibitem[\protect\citeauthoryear{{Field}, {Goldsmith} \& {Habing}}{{Field}
  et~al.}{1969}]{Field69}
{Field} G.~B.,  {Goldsmith} D.~W.,    {Habing} H.~J.,  1969, \apjl, 155, L149+

\bibitem[\protect\citeauthoryear{{Franco}, {Shore} \& {Tenorio-Tagle}}{{Franco}
  et~al.}{1994}]{Franco94}
{Franco} J.,  {Shore} S.~N.,    {Tenorio-Tagle} G.,  1994, \apj, 436, 795

\bibitem[\protect\citeauthoryear{{Fryxell}, {Olson}, {Ricker}, {Timmes},
  {Zingale}, {Lamb}, {MacNeice}, {Rosner}, {Truran} \& {Tufo}}{{Fryxell}
  et~al.}{2000}]{FLASH00}
{Fryxell} B.,  {Olson} K.,  {Ricker} P.,  {Timmes} F.~X.,  {Zingale} M.,
  {Lamb} D.~Q.,  {MacNeice} P.,  {Rosner} R.,  {Truran} J.~W.,    {Tufo} H.,
  2000, \apjs, 131, 273

\bibitem[\protect\citeauthoryear{{Gazol}, {V{\'a}zquez-Semadeni} \&
  {Kim}}{{Gazol} et~al.}{2005}]{Gazol05}
{Gazol} A.,  {V{\'a}zquez-Semadeni} E.,    {Kim} J.,  2005, \apj, 630, 911

\bibitem[\protect\citeauthoryear{{Gazol}, {V{\'a}zquez-Semadeni},
  {S{\'a}nchez-Salcedo} \& {Scalo}}{{Gazol} et~al.}{2001}]{Gazol01}
{Gazol} A.,  {V{\'a}zquez-Semadeni} E.,  {S{\'a}nchez-Salcedo} F.~J.,
  {Scalo} J.,  2001, \apjl, 557, L121

\bibitem[\protect\citeauthoryear{{Goldsmith} \& {Li}}{{Goldsmith} \&
  {Li}}{2005}]{Goldsmith05}
{Goldsmith} P.~F.,  {Li} D.,  2005, \apj, 622, 938

\bibitem[\protect\citeauthoryear{{G{\'o}mez}, {V{\'a}zquez-Semadeni},
  {Shadmehri} \& {Ballesteros-Paredes}}{{G{\'o}mez} et~al.}{2007}]{Gomez07}
{G{\'o}mez} G.~C.,  {V{\'a}zquez-Semadeni} E.,  {Shadmehri} M.,
  {Ballesteros-Paredes} J.,  2007, \apj, 669, 1042

\bibitem[\protect\citeauthoryear{{Gressel}}{{Gressel}}{2009}]{Gressel09}
{Gressel} O.,  2009, ArXiv e-prints

\bibitem[\protect\citeauthoryear{{Hartmann}, {Ballesteros-Paredes} \&
  {Bergin}}{{Hartmann} et~al.}{2001}]{Hartmann01}
{Hartmann} L.,  {Ballesteros-Paredes} J.,    {Bergin} E.~A.,  2001, \apj, 562,
  852

\bibitem[\protect\citeauthoryear{{Hartmann} \& {Burkert}}{{Hartmann} \&
  {Burkert}}{2007}]{Hartmann07}
{Hartmann} L.,  {Burkert} A.,  2007, \apj, 654, 988

\bibitem[\protect\citeauthoryear{{Heiles} \& {Troland}}{{Heiles} \&
  {Troland}}{2005}]{Heiles05}
{Heiles} C.,  {Troland} T.~H.,  2005, \apj, 624, 773

\bibitem[\protect\citeauthoryear{{Heitsch}, {Burkert}, {Hartmann}, {Slyz} \&
  {Devriendt}}{{Heitsch} et~al.}{2005}]{Heitsch05}
{Heitsch} F.,  {Burkert} A.,  {Hartmann} L.~W.,  {Slyz} A.~D.,    {Devriendt}
  J.~E.~G.,  2005, \apjl, 633, L113

\bibitem[\protect\citeauthoryear{{Heitsch} \& {Hartmann}}{{Heitsch} \&
  {Hartmann}}{2008}]{Heitsch08c}
{Heitsch} F.,  {Hartmann} L.,  2008, \apj, 689, 290

\bibitem[\protect\citeauthoryear{{Heitsch}, {Hartmann}, {Slyz}, {Devriendt} \&
  {Burkert}}{{Heitsch} et~al.}{2008}]{Heitsch08a}
{Heitsch} F.,  {Hartmann} L.~W.,  {Slyz} A.~D.,  {Devriendt} J.~E.~G.,
  {Burkert} A.,  2008, \apj, 674, 316

\bibitem[\protect\citeauthoryear{{Hennebelle} \& {Audit}}{{Hennebelle} \&
  {Audit}}{2007}]{Hennebelle07c}
{Hennebelle} P.,  {Audit} E.,  2007, \aap, 465, 431

\bibitem[\protect\citeauthoryear{{Hennebelle}, {Audit} \&
  {Miville-Desch{\^e}nes}}{{Hennebelle} et~al.}{2007}]{Hennebelle07}
{Hennebelle} P.,  {Audit} E.,    {Miville-Desch{\^e}nes} M.-A.,  2007, \aap,
  465, 445

\bibitem[\protect\citeauthoryear{{Hennebelle}, {Banerjee},
  {V{\'a}zquez-Semadeni}, {Klessen} \& {Audit}}{{Hennebelle}
  et~al.}{2008}]{Hennebelle08b}
{Hennebelle} P.,  {Banerjee} R.,  {V{\'a}zquez-Semadeni} E.,  {Klessen} R.~S.,
    {Audit} E.,  2008, \aap, 486, L43

\bibitem[\protect\citeauthoryear{{Hennebelle} \& {Inutsuka}}{{Hennebelle} \&
  {Inutsuka}}{2006}]{Hennebelle06}
{Hennebelle} P.,  {Inutsuka} S.-i.,  2006, \apj, 647, 404

\bibitem[\protect\citeauthoryear{{Hennebelle} \& {P{\'e}rault}}{{Hennebelle} \&
  {P{\'e}rault}}{1999}]{Hennebelle99}
{Hennebelle} P.,  {P{\'e}rault} M.,  1999, \aap, 351, 309

\bibitem[\protect\citeauthoryear{{Hennebelle} \& {P{\'e}rault}}{{Hennebelle} \&
  {P{\'e}rault}}{2000}]{Hennebelle00}
{Hennebelle} P.,  {P{\'e}rault} M.,  2000, \aap, 359, 1124

\bibitem[\protect\citeauthoryear{{Inoue} \& {Inutsuka}}{{Inoue} \&
  {Inutsuka}}{2008}]{Inoue08}
{Inoue} T.,  {Inutsuka} S.-i.,  2008, \apj, 687, 303

\bibitem[\protect\citeauthoryear{{Inoue}, {Inutsuka} \& {Koyama}}{{Inoue}
  et~al.}{2007}]{Inoue07}
{Inoue} T.,  {Inutsuka} S.-i.,    {Koyama} H.,  2007, \apjl, 658, L99

\bibitem[\protect\citeauthoryear{{Jappsen}, {Klessen}, {Larson}, {Li} \& {Mac
  Low}}{{Jappsen} et~al.}{2005}]{Jappsen05}
{Jappsen} A.-K.,  {Klessen} R.~S.,  {Larson} R.~B.,  {Li} Y.,    {Mac Low}
  M.-M.,  2005, \aap, 435, 611

\bibitem[\protect\citeauthoryear{{Klessen}}{{Klessen}}{2001}]{Klessen01}
{Klessen} R.~S.,  2001, \apj, 556, 837

\bibitem[\protect\citeauthoryear{{Klessen}, {Heitsch} \& {Mac Low}}{{Klessen}
  et~al.}{2000}]{Klessen00b}
{Klessen} R.~S.,  {Heitsch} F.,    {Mac Low} M.-M.,  2000, \apj, 535, 887

\bibitem[\protect\citeauthoryear{{Klessen} \& {Lin}}{{Klessen} \&
  {Lin}}{2003}]{Klessen03b}
{Klessen} R.~S.,  {Lin} D.~N.,  2003, \pre, 67, 046311

\bibitem[\protect\citeauthoryear{{Koyama} \& {Inutsuka}}{{Koyama} \&
  {Inutsuka}}{2000}]{Koyama00}
{Koyama} H.,  {Inutsuka} S.-I.,  2000, \apj, 532, 980

\bibitem[\protect\citeauthoryear{{Koyama} \& {Inutsuka}}{{Koyama} \&
  {Inutsuka}}{2002}]{Koyama02}
{Koyama} H.,  {Inutsuka} S.-i.,  2002, \apjl, 564, L97

\bibitem[\protect\citeauthoryear{{Koyama} \& {Inutsuka}}{{Koyama} \&
  {Inutsuka}}{2004}]{Koyama04}
{Koyama} H.,  {Inutsuka} S.-i.,  2004, \apjl, 602, L25

\bibitem[\protect\citeauthoryear{{Kwan} \& {Valdes}}{{Kwan} \&
  {Valdes}}{1983}]{Kwan83}
{Kwan} J.,  {Valdes} F.,  1983, \apj, 271, 604

\bibitem[\protect\citeauthoryear{{Mac Low} \& {Klessen}}{{Mac Low} \&
  {Klessen}}{2004}]{MacLow04}
{Mac Low} M.-M.,  {Klessen} R.~S.,  2004, Reviews of Modern Physics, 76, 125

\bibitem[\protect\citeauthoryear{{McKee} \& {Ostriker}}{{McKee} \&
  {Ostriker}}{2007}]{McKee07}
{McKee} C.~F.,  {Ostriker} E.~C.,  2007, \araa, 45, 565

\bibitem[\protect\citeauthoryear{{Mestel}}{{Mestel}}{1985}]{Mestel85}
{Mestel} L.,  1985, in {Black} D.~C.,  {Matthews} M.~S.,  eds, Protostars and
  Planets II {Magnetic fields}.
pp 320--339

\bibitem[\protect\citeauthoryear{{Mouschovias} \& {Spitzer} Jr.}{{Mouschovias}
  \& {Spitzer}}{1976}]{Mouschovias76}
{Mouschovias} T.~C.,  {Spitzer} Jr. L.,  1976, \apj, 210, 326

\bibitem[\protect\citeauthoryear{{Passot}, {Vazquez-Semadeni} \&
  {Pouquet}}{{Passot} et~al.}{1995}]{Passot95}
{Passot} T.,  {Vazquez-Semadeni} E.,    {Pouquet} A.,  1995, \apj, 455, 536

\bibitem[\protect\citeauthoryear{{S{\'a}nchez-Salcedo}, {V{\'a}zquez-Semadeni}
  \& {Gazol}}{{S{\'a}nchez-Salcedo} et~al.}{2002}]{Sanchez02}
{S{\'a}nchez-Salcedo} F.~J.,  {V{\'a}zquez-Semadeni} E.,    {Gazol} A.,  2002,
  \apj, 577, 768

\bibitem[\protect\citeauthoryear{{Shu}, {Allen}, {Lizano} \& {Galli}}{{Shu}
  et~al.}{2007}]{Shu07}
{Shu} F.~H.,  {Allen} R.~J.,  {Lizano} S.,    {Galli} D.,  2007, \apjl, 662,
  L75

\bibitem[\protect\citeauthoryear{{Spitzer}}{{Spitzer}}{1978}]{SpitzerBook78}
{Spitzer} L.,  1978, {Physical processes in the interstellar medium}.
New York Wiley-Interscience, 1978.~333 p.

\bibitem[\protect\citeauthoryear{{Truelove}, {Klein}, {McKee}, {Holliman},
  {Howell} \& {Greenough}}{{Truelove} et~al.}{1997}]{Truelove97}
{Truelove} J.~K.,  {Klein} R.~I.,  {McKee} C.~F.,  {Holliman} J.~H.,  {Howell}
  L.~H.,    {Greenough} J.~A.,  1997, \apjl, 489, L179+

\bibitem[\protect\citeauthoryear{{V{\'a}zquez-Semadeni}}{{V{\'a}zquez-Semadeni%
}}{2007}]{Vazquez07b}
{V{\'a}zquez-Semadeni} E.,  2007, in {Elmegreen} B.~G.,  {Palous} J.,  eds, IAU
  Symposium Vol.~237 of IAU Symposium, {Molecular cloud turbulence and the star
  formation efficiency: enlarging the scope}.
pp 292--299

\bibitem[\protect\citeauthoryear{{V{\'a}zquez-Semadeni}, {Ballesteros-Paredes}
  \& {Klessen}}{{V{\'a}zquez-Semadeni} et~al.}{2003}]{Vazquez03a}
{V{\'a}zquez-Semadeni} E.,  {Ballesteros-Paredes} J.,    {Klessen} R.~S.,
  2003, \apjl, 585, L131

\bibitem[\protect\citeauthoryear{{V{\'a}zquez-Semadeni}, {Ballesteros-Paredes},
  {Klessen} \& {Jappsen}}{{V{\'a}zquez-Semadeni} et~al.}{2008}]{Vazquez08}
{V{\'a}zquez-Semadeni} E.,  {Ballesteros-Paredes} J.,  {Klessen} R.~S.,
  {Jappsen} A.~K.,  2008, in {Beuther} H.,  {Linz} H.,   {Henning} T.,  eds,
  Astronomical Society of the Pacific Conference Series Vol.~387 of
  Astronomical Society of the Pacific Conference Series, {Massive Star-Forming
  Regions: Turbulent Support or Global Collapse?}.
pp 240--+

\bibitem[\protect\citeauthoryear{{Vazquez-Semadeni}, {Ballesteros-Paredes} \&
  {Rodriguez}}{{Vazquez-Semadeni} et~al.}{1997}]{Vazquez97}
{Vazquez-Semadeni} E.,  {Ballesteros-Paredes} J.,    {Rodriguez} L.~F.,  1997,
  \apj, 474, 292

\bibitem[\protect\citeauthoryear{{V{\'a}zquez-Semadeni}, {Gazol}, {Passot} \&
  {et al.}}{{V{\'a}zquez-Semadeni} et~al.}{2003}]{Vazquez03b}
{V{\'a}zquez-Semadeni} E.,  {Gazol} A.,  {Passot} T.,    {et al.} 2003, in
  {Falgarone} E.,  {Passot} T.,  eds, Turbulence and Magnetic Fields in
  Astrophysics Vol.~614 of Lecture Notes in Physics, Berlin Springer Verlag,
  {Thermal Instability and Magnetic Pressure in the Turbulent Interstellar
  Medium}.
pp 213--251

\bibitem[\protect\citeauthoryear{{V{\'a}zquez-Semadeni}, {Gazol} \&
  {Scalo}}{{V{\'a}zquez-Semadeni} et~al.}{2000}]{Vazquez00}
{V{\'a}zquez-Semadeni} E.,  {Gazol} A.,    {Scalo} J.,  2000, \apj, 540, 271

\bibitem[\protect\citeauthoryear{{V{\'a}zquez-Semadeni}, {G{\'o}mez},
  {Jappsen}, {Ballesteros-Paredes}, {Gonz{\'a}lez} \&
  {Klessen}}{{V{\'a}zquez-Semadeni} et~al.}{2007}]{Vazquez07}
{V{\'a}zquez-Semadeni} E.,  {G{\'o}mez} G.~C.,  {Jappsen} A.~K.,
  {Ballesteros-Paredes} J.,  {Gonz{\'a}lez} R.~F.,    {Klessen} R.~S.,  2007,
  \apj, 657, 870

\bibitem[\protect\citeauthoryear{{V{\'a}zquez-Semadeni}, {G\'omez}, {Jappsen},
  {Ballesteros-Paredes} \& {Klessen}}{{V{\'a}zquez-Semadeni}
  et~al.}{2009}]{Vazquez09}
{V{\'a}zquez-Semadeni} E.,  {G\'omez} G.~C.,  {Jappsen} A.~K.,
  {Ballesteros-Paredes} J.,    {Klessen} R.~S.,  2009, arXiv:astro-ph

\bibitem[\protect\citeauthoryear{{V{\'a}zquez-Semadeni}, {Ryu}, {Passot},
  {Gonz{\'a}lez} \& {Gazol}}{{V{\'a}zquez-Semadeni} et~al.}{2006}]{Vazquez06}
{V{\'a}zquez-Semadeni} E.,  {Ryu} D.,  {Passot} T.,  {Gonz{\'a}lez} R.~F.,
  {Gazol} A.,  2006, \apj, 643, 245

\bibitem[\protect\citeauthoryear{{Zeldovich} \& {Pikel'Ner}}{{Zeldovich} \&
  {Pikel'Ner}}{1969}]{Zeldovich69}
{Zeldovich} Y.~B.,  {Pikel'Ner} S.~B.,  1969, Soviet Journal of Experimental
  and Theoretical Physics, 29, 170

\bibitem[\protect\citeauthoryear{{Zinnecker} \& {Yorke}}{{Zinnecker} \&
  {Yorke}}{2007}]{ZinneckerYorke07}
{Zinnecker} H.,  {Yorke} H.~W.,  2007, \araa, 45, 481

\end{thebibliography}
